\documentclass[11pt,a4paper]{article}
\pdfoutput=1

\usepackage{jcappub}
\usepackage[T1]{fontenc}
\usepackage{amsmath}
\usepackage{amssymb}
\usepackage{amsthm}
\usepackage{amsopn}
\usepackage{verbatim}
\usepackage{graphicx}
\usepackage{psfrag}

\newcommand{\ltsim}{\mathrel{\raise.3ex\hbox{$<$\kern-.75em\lower1ex\hbox{$\sim$}}}}
\newcommand{\gtsim}{\mathrel{\raise.3ex\hbox{$>$\kern-.75em\lower1ex\hbox{$\sim$}}}}

\newcommand{\prd}{PRD}
\newcommand{\apj}{ApJ}
\newcommand{\apjs}{ApJS}
\newcommand{\jcap}{JCAP}
\newcommand{\mnras}{MNRAS}
\newcommand{\apjl}{ApJL}
\newcommand{\physrep}{PhysRep}
\newcommand{\procspie}{ProcSpie}

\notoc

\begin{document}

\title{The information content of cosmic microwave background anisotropies}

\author[a]{Douglas Scott,\note{Corresponding author.}}
\author[a]{Dagoberto Contreras,}
\author[a]{Ali Narimani,}
\author[b]{Yin-Zhe Ma}

\affiliation[a]{Department of Physics and Astronomy,\\
University of British Columbia,\\
Vancouver, BC, Canada\ V6T1Z1}

\affiliation[b]{Astrophysics and Cosmology Research Unit,\\
School of Chemistry and Physics,\\
University of KwaZulu-Natal,\\
Durban 4041, South Africa}

\emailAdd{dscott@phas.ubc.ca}
\emailAdd{dagocont@phas.ubc.ca}
\emailAdd{anariman@phas.ubc.ca}
\emailAdd{ma@ukzn.ac.za}

\abstract{
The cosmic microwave background (CMB) contains perturbations that are close
to Gaussian and isotropic.  This means that its information content,
in the sense of
the ability to constrain cosmological models, is closely related to the number
of modes probed in CMB power spectra.  Rather than making forecasts for
specific experimental setups, here we take a more pedagogical approach and
ask how much information we can extract from the CMB if we are only limited
by sample variance.  We show that, compared with temperature measurements,
the addition of $E$-mode polarization doubles
the number of modes available out to a fixed maximum multipole, provided
that all of the $TT$, $TE$, and $EE$ power spectra are measured.  However,
the situation in terms of constraints on particular parameters is more
complicated, as we explain and illustrate graphically.
We also discuss the enhancements in information
that can come from adding $B$-mode polarization and gravitational lensing.
We show how well one could ever determine the basic cosmological parameters
from CMB data compared with what has been achieved with {\it Planck},
which has already probed a substantial fraction of the $TT$ information.
Lastly, we look at constraints on neutrino mass as a specific example of how
lensing information improves future prospects beyond the current
6-parameter model.}


\maketitle
\hfil
\flushbottom

\section{Introduction}
\label{sec:Intro}
{\it Planck\/} \cite{PlanckI}, the {\it Wilkinson Microwave Anisotropy Probe\/}
(WMAP \cite{WMAP}), the Atacama Cosmology Telescope (ACT \cite{ACT}),
the South Pole Experiment (SPT \cite{SPT})
and other cosmic microwave background (CMB) experiments have
measured the CMB with high sensitivity, covering angular scales from
essentially the whole sky down to the arcminute regime \cite{ScottSmoot}.
It is well known that the precision
with which these measurements are made (together with an understanding of the
physics generating the anisotropies) allows us to place very tight
constraints on cosmological parameters.  A large number of studies have
focused on predicting how well the parameters can be constrained using
existing or future CMB data \cite{Jungman96,BET97,ZSS97,EB99,Rocha04,
Perotto06,Burigana10,Foreman11,Wu14,Galli14}.
In this paper we want to take a rather more global view, and ask just how much
constraining power there is to be mined, or in other words, how much
cosmological information there is to extract from the CMB anisotropies.
This will lead us to address questions like:
\begin{itemize}
\item How is the overall information content related to the number of CMB
modes measured?
\item What does polarization add to temperature information?
\item How do power spectrum measurements relate to constraints on
parameters?
\item Is the information finite, and how far have we progressed towards the
goal of measuring all that is available?
\end{itemize}

The basic 2015 {\it Planck\/} data set (including fairly conservative masking
of the sky, as well as fitting of foreground signals) gives a temperature power
spectrum that is measured to approximately $750\,\sigma$ and
polarization-related power spectra that are measured to around $280\,\sigma$
\cite{PlanckI}.  We might therefore naively expect that there is about
$800\,\sigma$
worth of constraints to be shared out among the cosmological parameters.
However, a check of the {\it Planck}-derived constraints shows that this same
data set yields a value for $100\theta_\ast$ (a parameterization of the
ratio of the sound horizon to the last-scattering surface distance) that is
$1.04103\pm0.00046$ \cite{PlanckXIII}, which corresponds to an almost
$2300\,\sigma$ measurement of $\theta_\ast$.  At the same time, the
constraints on the other five parameters in the usual set give a quadrature
sum of about $120\,\sigma$ (which is fairly negligible compared to the
$\theta_\ast$ constraint).

The original motivation for this paper was to ask ``how is it that an
$800\,\sigma$ measurement of anisotropy power leads to a roughly $2300\,\sigma$
combined constraint on cosmological parameters?''  In attempting to answer
this question, we hope to illuminate some issues concerning
cosmological parameter constraints in general, and how they might relate
to experimental design in the future.

In this paper we focus on the conventional cosmological-constant-dominated
cold dark matter model, $\Lambda$CDM, and a standard 6-parameter set of
cosmological parameters: $A_{\rm s}$, the amplitude of the initial power
spectrum; $n_{\rm s}$, the power-law slope of the initial conditions;
$\Omega_{\rm b}h^2$, the baryonic density;
$\Omega_{\rm c}h^2$, the cold dark matter density;
$\theta_\ast$, which we have already defined;
and $\tau$, the optical depth to reionization.
Here $h$ is the Hubble parameter today, $H_0$, in units of
$100\,{\rm km}\,{\rm s}^{-1}\,{\rm Mpc}^{-1}$.  Within this model we use
the code {\tt CAMB} \cite{Lewis2000} to calculate CMB power spectra.

\section{CMB anisotropy information}
\label{sec:CMB}
The word ``information'' has many different meanings.  Here, we use the
word to mean the strength of our ability to constrain cosmologies.
The total amount of information available for cosmological surveys is related
to the number of observable modes \cite{Loeb12,MaScott}.  This situation has
been described in many papers related to measuring the 3-dimensional power
spectrum in order to constrain cosmological parameters
\cite{Seljak98,Padmanabhan01,Rimes05,McDonald09,Carron15}.

The situation for CMB temperature anisotropies is simpler, since it
only involves assessing the information content of a purely 2-dimensional sky.
On the other hand, as we shall see the relationship between the modes and the
constraints on cosmological parameters is non-trivial.

Let us start by recalling that the temperature field $T(\theta,\phi)$ on the
sky is usually expanded in terms of spherical harmonics, i.e.,
\begin{equation}
T(\theta,\phi) = \sum_{\ell=2}^\infty \sum_{m=-\ell}^{+\ell}
 a_{\ell m} Y_{\ell m}(\theta,\phi),
\end{equation}
where we have removed the monopole (average CMB temperature) and dipole
(which is dominated by our local velocity).
Then, provided one goes to sufficiently high multipoles, and ignoring
effects of beams and masking, one can use the set of $a_{\ell m}$s as an
alternative representation of the pixels in the map.  The power spectrum
$C_\ell$ is the expectation value of the variance of the $a_{\ell m}$s as a
function of $\ell$, with each $m$ being equivalent since there are no
cosmologically preferred directions.

If the perturbations are Gaussian then each sky is a realization of this power
spectrum. The scatter among these realizations is known as ``cosmic variance.''
The cosmic variance in estimates of the $C_{\ell}$s on the full sky is
\begin{equation}
 \Delta C_{\ell}=\sqrt{\frac{2}{(2\ell+1)}}C_{\ell}
\label{eq:CV}
\end{equation}
(e.g., Ref.~\cite{AbbottWise}).
The factor of 2 here is because this is effectively the ``variance of the
variance,'' and for a Gaussian distribution that is twice the variance.  The
factor of $(2\ell+1)$ is the number of $m$ modes for each $\ell$ and if only
a fraction $f_{\rm sky}$ is observed, then the approximate effect is to
increase the uncertainty in $C_\ell$ so that the ``sample variance'' is
$f_{\rm sky}^{-1}$ larger \cite{ScottSW}.  A more realistic expression
can also be
written that includes the instrumental noise and beam, as described in
Ref.~\cite{Knox}.

For the simple case of an all-sky, noise-free experiment, which
measures multipoles perfectly up to $\ell_{\rm max}$, the
total square of the signal-to-noise ratio (using Eq.~\ref{eq:CV}) is
\begin{eqnarray}
(S/N)^2 &\equiv&
 \sum^{\ell_{\rm max}}_{\ell=2}\left(C_{\ell}/\Delta C_{\ell}\right)^{2}
 = \frac{1}{2} \sum^{\ell_{\rm max}}_{\ell=2} (2\ell+1) \nonumber\\
 &=& \frac{1}{2} \Big[\ell_{\rm max}(\ell_{\rm max}+2)-3\Big].
 \label{eq:SN_Cl}
\end{eqnarray}
Note that this calculation is {\it exactly half\/} the total number of
modes, i.e., $\sum (2\ell+1)$.  This means that in terms of
constraints on the power spectrum, each $a_{\ell m}$ mode contributes
$1/2$ to the square of the total signal-to-noise ratio (SNR).
In other words, estimating the information in the power spectrum is
effectively the same thing as counting modes.

To be clear, we are distinguishing here between trivial
information that tells us about the particular realization of our
Universe (which we can continue to measure as precisely as we
wish) and the more valuable information we can extract from our
Hubble patch, which gives us constraints on the background
cosmological model.  The fact that the CMB sky is remarkably close to Gaussian
allows us to reduce the information contained in the individual $a_{\ell m}$s
(or equivalently in the particular hot and cold spots on our sky) to estimates
of the power at each $\ell$ (or equivalently the variance among pixels as a
function of angular separation).  For Gaussian skies, the
amount of information in the power spectrum is directly proportional to the
number of independent modes that can be measured.

Certainly one {\it could\/} regard all the data in a map as being
``information,'' i.e., the fact that there is a CMB hot spot in a particular
direction is of some consequence, just as it matters that we live in the Milky
Way galaxy, rather than M31.  However, we discount these particulars about our
realization, since they do not tell us about our overall cosmological model.
So here, when we say ``information,'' we are referring to the constraining
power for cosmological models, or more specifically the signal-to-noise
ratio.  Clearly there is a relationship between this ``information'' and the
amount of computer memory required to store the related data, i.e., the number
of bits needed.  In a more formal information theoretical sense, the number
of bits required corresponds to the base~2 logarithm of what
we are defining as information (e.g., Ref.~\cite{Grandis15}) -- but here we are
talking about signal-to-noise ratio for power spectra, since that is where
parameter constraints come from.

For the case of the CMB each $a_{\ell m}$ is a random number coming from a
Gaussian distribution, with mean zero and variance $C_\ell$.  To
estimate the $C_\ell$ from a set of observed $a_{\ell m}$s, it is
sufficient to have only a very few bits of information for each
$a_{\ell m}$, since we only need each $a_{\ell m}$ to help us obtain an
estimate of the variance.  Hence the amount of cosmological information --
i.e., the ability to eventually constrain parameters --
is determined by the number of measurable modes times a (roughly) constant (but
parameterization dependent) numerical factor.
However, for a particular parameter it may be that some modes are more
important than others; to understand how the information of power spectra maps
onto the information of the cosmological model, we can perform a more rigorous calculation by
considering the Fisher matrix, as we do in the following sections.

\section{CMB Fisher information}
\label{sec:Fisher}
The previous section considered CMB temperature anisotropies only, but since
the CMB sky can be linearly polarized, there exists additional information
in each pixel of a CMB map.  If an experiment can measure the $Q$ and $U$
Stokes parameters (in addition to $T$), then in principle there are two
additional pieces of information for each pixel.  The most useful way of
describing these additional degrees of freedom is through a geometrical
approach, defining a divergence-like combination, usually called ``$E$,'' and a
curl-like combination, usually called ``$B$'' \cite{ZS97,KKS97,HW97}.
$T$, $E$, and $B$ fields can be
used to determine auto- and cross-power spectra, and with parity considerations
(at least for cosmological signals) making $TB$ and $EB$ uncorrelated, we are
left with four CMB power spectra from which we can constrain parameters,
namely $C_\ell^{TT}$, $C_\ell^{TE}$, $C_\ell^{EE}$, and $C_\ell^{BB}$.

The $T$- and $E$-mode maps have now been well measured
\cite{Hinshaw12,Story13,Das14,Naess14,PlanckXI,Crites15,KB15},
but estimates of $B$-modes are still in their infancy \cite{BKP}.
Moreover, even when primordial $B$-modes become detectable, we expect
them to be small, and hence one will need an experiment with an
entirely different sensitivity range to probe the information contained in
those modes.  For these reasons we will neglect $B$-modes in most of the
discussion of this paper.  However, the $B$-modes caused by the effects of
gravitational lensing have now been detected by several experiments
\cite{Hanson13,POLARBEAR,Keisler,PlanckXV,KB15,vanEngelen15}, and we will
discuss this later in Section~\ref{sec:other}.

Focusing only on $T$ and $E$, an important fact is that the two fields on
the sky are {\it not\/} independent, but contain correlations, which can
be measured through the cross-power spectrum $C_{\ell}^{TE}$.
Because of this, it may not be entirely obvious how much additional information
is provided by measurements of CMB polarization -- does a measurement of $T$
and $E$ provide twice as much information as provided by each of them alone?
We will answer this question in the following subsections.

\subsection{The Fisher matrix}
The Fisher matrix gives a powerful formalism for describing the information
content coming from observables in terms of underlying parameters (see
e.g., Ref.~\cite{TTH97}).  Under the assumption of Gaussian
perturbations and with negligible instrumental noise, the Fisher
information matrix for CMB temperature and polarization
anisotropies is \cite{Eisenstein98}
\begin{equation}
F_{ij}=\sum_{\ell}
 \sum\limits_{XY}\frac{\partial C^X_\ell}{\partial p_{i}}
 (\mathbb{C}_{\ell})_{XY}^{-1}\frac{\partial C^Y_\ell}{\partial p_{j}},
\label{eq:Fisher}
\end{equation}
where $C^X_\ell$ and $C^Y_\ell$ are the power in the $\ell$th
multipole for $X, Y = T$, $E$, or $C$ (temperature, $E$-mode polarization, and
$TE$ correlation, respectively), and the $p_i$ are cosmological parameters.
Here we are ignoring the $B$-modes, as already explained;
however, in principle one could easily extend Eq.~(\ref{eq:Fisher}) to
include them.

We now define the vector $\vec{x}_{\ell}$ as
\begin{equation}
\label{eq:vector}
\vec{x}_{\ell}=\left(
 \begin{array}{c}
  C_{\ell}^{TT} \\
  C_{\ell}^{EE} \\
  C_{\ell}^{TE}
 \end{array}
\right) ,
\end{equation}
and $\mathbb{C}_{\ell}$ as a covariance matrix
\begin{equation}
 \mathbb{C}_{\ell}=\left(
 \begin{array}{ccc}
  (\mathbb{C}_{\ell})_{TT} & (\mathbb{C}_{\ell})_{TE} &
  (\mathbb{C}_{\ell})_{TC} \\
  (\mathbb{C}_{\ell})_{TE} & (\mathbb{C}_{\ell})_{EE} &
  (\mathbb{C}_{\ell})_{EC} \\
  (\mathbb{C}_{\ell})_{TC} & (\mathbb{C}_{\ell})_{EC} &
  (\mathbb{C}_{\ell})_{CC}
 \end{array}
\right) .  \label{eq:cov}
\end{equation}
We can then formulate Eq.~(\ref{eq:Fisher}) into a matrix product as
\begin{equation}
 F_{ij}=\sum_{\ell}\frac{\partial \vec{x}_{\ell}^{\sf T}}{\partial p_{i}}
 (\mathbb{C}_{\ell})^{-1}\frac{\partial \vec{x}_{\ell}}{\partial p_{j}},
\end{equation}
where the $(\mathbb{C}_{\ell})_{XY}$ entries for the noise-free case are
\begin{eqnarray}
 (\mathbb{C}_{\ell})_{TT} &=&\frac{2}{(2\ell +1)f_{\rm sky}}
 (C_{\ell}^{TT})^{2}, \nonumber\\
 (\mathbb{C}_{\ell})_{EE} &=&\frac{2}{(2\ell +1)f_{\rm sky}}
 (C_{\ell}^{EE})^{2}, \nonumber\\
 (\mathbb{C}_{\ell})_{TE} &=&\frac{2}{(2\ell +1)f_{\rm sky}}
 (C_{\ell}^{TE})^{2}, \nonumber\\
 (\mathbb{C}_{\ell})_{TC} &=&\frac{2}{(2\ell +1)f_{\rm sky}}
 (C_{\ell}^{TT}C_{\ell}^{TE}), \nonumber\\
 (\mathbb{C}_{\ell})_{EC} &=&\frac{2}{(2\ell +1)f_{\rm sky}}
 (C_{\ell}^{EE}C_{\ell}^{TE}), \nonumber\\
 (\mathbb{C}_{\ell})_{CC} &=&\frac{1}{(2\ell +1)f_{\rm sky}}
 \Big[C_{\ell}^{EE}C_{\ell}^{TT}+(C_{\ell}^{TE})^{2}\Big],
\end{eqnarray}
following Ref.~\cite{Eisenstein98}. For convenience we will define $\mathcal{N}
\equiv f_{\rm sky} (2\ell+1)/2$ and $r_{\ell} \equiv
C^{TE}_{\ell}/\sqrt{C^{TT}_{\ell}C^{EE}_{\ell}}$, which is the correlation
coefficient between $T$ and $E$ (see appendix A6 in Ref.~\cite{PlanckXI}).

The Cramer-Rao bound states that we can assign the $1\,\sigma$
statistical uncertainties to be $\sigma_i = \sqrt{(F^{-1})_{ii}}$, this gives
the smallest possible errors achievable.  Now we will
consider the simplest situation, where there is only one cosmological parameter
to determine. This scenario, though simple, will explain the effects of
including polarization along with temperature data, given that the
{\it maps\/} are
correlated.  In Section~\ref{sec:Params} we will consider larger parameter sets
which will explain the effects of correlations between {\it parameters\/}.

\begin{figure*}[htb!]
\centering
\includegraphics[width=\textwidth]{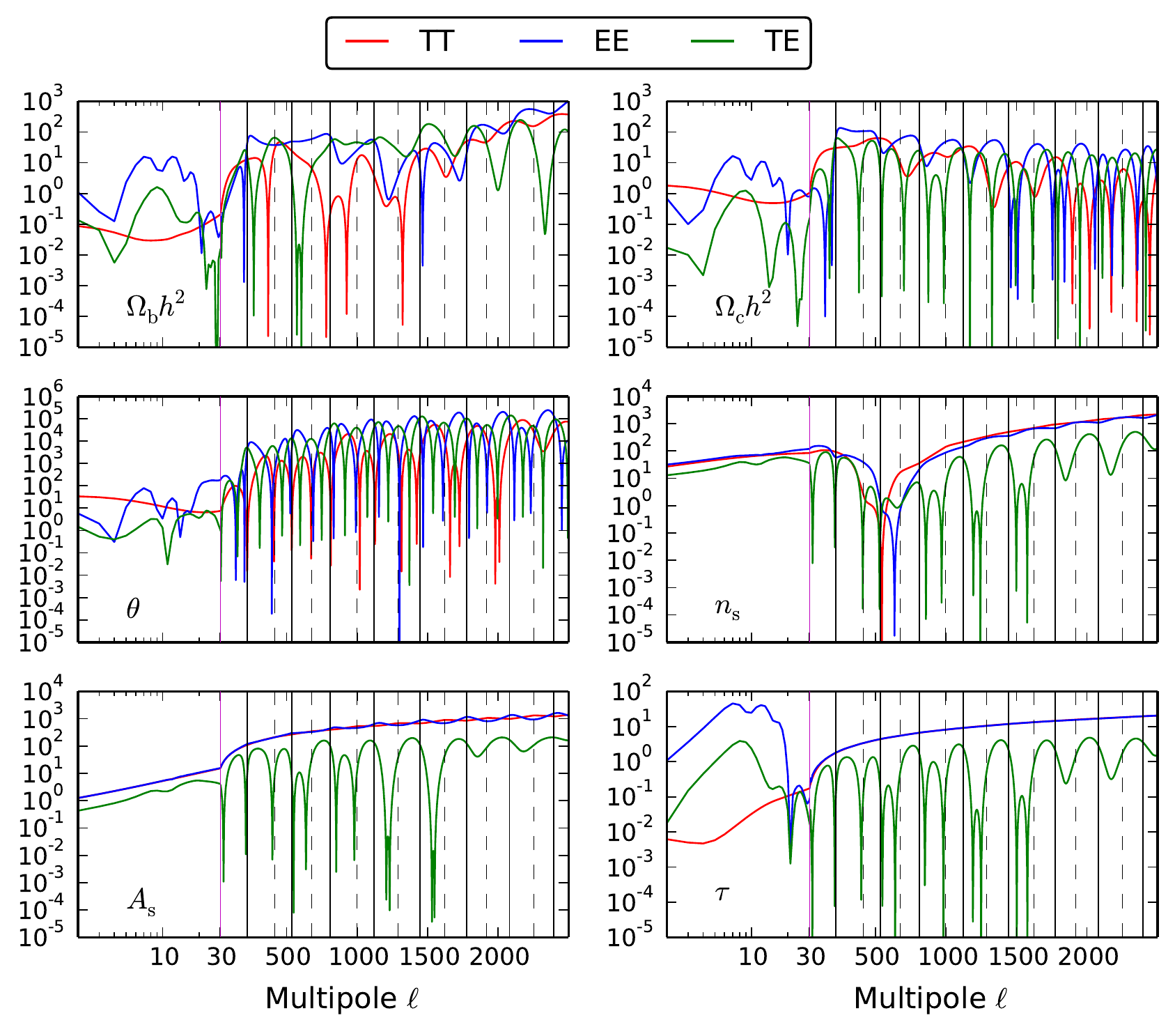}
\caption {Square of the signal-to-noise ratio on an $\ell$-by-$\ell$ basis for
the usual six parameters of the standard $\Lambda$CDM model, using the best-fit
{\it Planck\/} cosmology as the fiducial model.  The curves are for a
hypothetical noise-free, $f_{\rm sky}=0.5$ experiement.  The results for the
$TT$ spectrum (see Eq.~\ref{eq:ttsignaltonoise}) are in red,
$EE$ (Eq.~\ref{eq:eesignaltonoise}) in blue,
and $TE$ (Eq.~\ref{eq:tandesignaltonoise}) in green.  To highlight the
low-$\ell$ region, we have plotted the first 30 multipoles using a logarithmic
$x$-axis (and drawn a vertical line to separate the log and linear parts).
For reference, solid vertical lines mark the positions of peaks
in the $TT$ power spectrum (approximately the same positions as troughs in
$EE$) and dashed vertical lines are troughs in $TT$ (or peaks in $EE$);
the numerical values of the peaks and troughs are taken from
Ref.~\cite{PlanckXI}.  Note the different ranges plotted on the $y$-axis of
each panel.}
\label{fig:SNRparams}
\end{figure*}

\subsection{\textit{TT} only}
Let us first consider the case in which only the CMB temperature is mapped,
i.e., we have not measured $Q$ and $U$ in this scenario.
If we focus on a single parameter $p$, then the SNR can be written as
\begin{equation}
\left( \frac{p}{\Delta p}\right) = pF_{pp}^{1/2},
\label{eq:parameternoise}
\end{equation}
where we sum over all the multipoles $\ell$.
For the temperature power spectrum alone, then the covariance matrix
just has a single entry and we can write the squared SNR at each multipole as
\begin{equation}
  \left(\frac{S^2}{N^2}\right)^{TT}_{\ell} = \mathcal{N}
  \left(\frac{d\ln{C^{TT}_{\ell}}}{d\ln p}\right)^2.
  \label{eq:ttsignaltonoise}
\end{equation}

If we have a simple scaling parameter (similar to $A_{\rm s}$) then
$\partial C^{X}_{\ell}/\partial p=C^{X}_{\ell}/ p$ and the logarithmic
derivative is unity, so we just have
\begin{eqnarray}
\left( \frac{p}{\Delta p}\right)^{TT}  &=& \left(\sum_{\ell}
\mathcal{N}\right)^{1/2}  \notag \\
& = & \left( f_{\rm sky} \sum_\ell \left[\ell + \frac{1}{2}\right] \right)^{1/2} \\
& \simeq &\frac{1}{\sqrt{2}}f_{\rm sky}^{1/2}\ell_{\max}.
\end{eqnarray}
Hence we see that in the simple case of a single parameter that is proportional
to the amplitude of the power spectrum, the constraining power on this
parameter from the $TT$ power spectrum is the same as the mode counting
presented in the previous section.

Things are more complicated when we consider general parameters.
Figure~\ref{fig:SNRparams} shows the squared SNR per multipole for the
six parameters of the standard $\Lambda$CDM model following
Eq.~(\ref{eq:ttsignaltonoise}), i.e., with all other parameters held
fixed.\footnote{Gaussianity is not a valid assumption for the low
multipoles, and hence the low-$\ell$ part of the plots should be taken
as approximate estimates only.} The red curves in each
panel show the $TT$-only case.  We can see many details from this
curve, illustrating how different multipole ranges affect constraints on
each parameter.

Variations due to the parameter $A_{\rm s}$ are similar to those we have
described for a ``scaling parameter,'' but not quite the same.
If we looked at the $TT$ spectrum without the effects
of gravitational lensing, then we would find that the spectrum scales
exactly like $A_{\rm s}e^{-2\tau}$, and hence this is the
scaling parameter we referred to in Eq.~(\ref{eq:ttsignaltonoise}).
Since for Fig.~\ref{fig:SNRparams} all of the parameters are held fixed
except each one individually, then
both $A_{\rm s}$ and $e^{-2\tau}$ serve effectively
as scaling parameters in the sense of Eq.~(\ref{eq:ttsignaltonoise}).
Because of this, the curve for
$\tau$ looks like the one for $A_{\rm s}$, multiplied by $4\tau^2$
(since $d\ln{C^{TT}_{\ell}}/d\ln\tau = -2\tau$),
except for an extra variation at the lowest $\ell$s.  When we include lensing
in the usual way (as we have done in Fig.~\ref{fig:SNRparams}) then there are
also small wiggles in the $A_{\rm s}$ curve, which come from the smoothing
effect of lensing on the peaks and troughs.

For the slope, $n_{\rm s}$, we see the effect of the ``pivot'' point at
$k=0.05\,{\rm Mpc}^{-1}$, which projects to $\ell\simeq550$ for $TT$; at
multipoles around this point there is no constraint on $n_{\rm s}$.
The variations for most of the other parameters reflect the structure of the
$C_\ell$s themselves.  For example, for $\theta_\ast$ the curve goes to zero
near the positions of peaks and troughs (because the gradient of $C_\ell$ is
zero there).

The $A_{\rm s}$ panel in Figure~\ref{fig:SNRparams} can serve as a guideline
for inferring the sensitivity of a $C_\ell$ to a particular parameter.
One can approximately define the dependency of $C_\ell$ to a generic parameter
$p$, by a polynomial as $C_\ell \propto p^{\alpha}$, with $\alpha = 1$ then
representing linear dependence, and we might speak of ``less than linear''
and ``non-linear'' as corresponding to $\alpha<1$ and $\alpha>1$, respectively.
Comparing each of the panels of Figure~\ref{fig:SNRparams} with $A_{\rm s}$,
shows that the set $\{\Omega_{\rm b}h^2, \Omega_{\rm c}h^2, \tau \}$ all have
a less than linear relation with the $C_{\ell}$s at $\ell \lesssim 30$,
while $n_{\rm s}$ exhibits a non-linear relationship. The parameter
$\theta_\ast$ is non-linear over almost the entire multipole range (note the
different $y$-axis range for this panel).  In the high-$\ell$
region, $\ell > 30$, all of the parameters (except $\theta_\ast$)
show a mildly less than linear relation, with oscillations close to zero at
some multipoles (and this is even true for $\theta_\ast$), while
$n_{\rm s}$ has a close to linear relation at $\ell \gtrsim 1500$.

We will discuss the blue and green ($EE$ and $TE$) curves in the following
subsections.

\subsection{\textit{EE} only}
Now consider the situation where only the $E$-modes are mapped, and
hence we only have access to the $EE$ power spectrum for constraining
cosmology.  Here the situation is clearly exactly the same as it was for
the $TT$-only case.

We have the squared SNR for a single parameter $p$ being
\begin{equation}
  \left(\frac{S^2}{N^2}\right)^{EE}_{\ell} = \mathcal{N}
  \left(\frac{d\ln{C^{EE}_{\ell}}}{d\ln p}\right)^2
  \label{eq:eesignaltonoise}
\end{equation}
and the total for a scaling parameter is
\begin{equation}
\left( \frac{p}{\Delta p}\right)^{EE}
 \simeq \frac{1}{\sqrt{2}}f_{\rm sky}^{1/2}\ell_{\max}.
\end{equation}
Again, the result is just what we expect from mode counting.  If we have
polarization data out to some $\ell_{\rm max}$, then it provides the
same constraints on a scaling parameter as having temperature
data out to the same $\ell_{\rm max}$.

The situation for more general parameters is presented by the blue curves in the
panels of Figure~\ref{fig:SNRparams}.  One can see several effects that are
similar to the $TT$ case.  The situation for $A_{\rm s}$ and $\tau$ are
essentially the same as for the $TT$ case, with dramatic improvement for $\tau$
at low $\ell$ because of the sensitivity of large-scale polarization to
reionization.  The constraining power for $n_{\rm s}$ also has a zero in the
$EE$ case, but the pivot $k_0$ projects to a slightly different $\ell$
(reflecting the slightly different scales with which polarization is sourced
compared to temperature).  For
the parameters $\theta_\ast$, $\Omega_{\rm b}h^2$, and $\Omega_{\rm c}h^2$
we see that the constraining power from $EE$ is generally higher than that
for $TT$ (as recently pointed out in Ref.~\cite{Galli14} and explained in the
next section).  This illustrates the improved parameter constraints from
polarization, essentially because of the sharper acoustic features in
polarization.

\subsection{\textit{TE} correlation only}
Now let us examine the case where we measure the $TE$ power spectrum
only.  This is clearly not a realistic situation, but evaluating it will
elucidate some interesting points.  The Fisher matrix again has a single entry,
coming from the term $(\mathbb{C}_{\ell})_{CC}$, and for a
single parameter $p$, we obtain
\begin{equation}
  \left(\frac{S^2}{N^2}\right)^{TE}_{\ell} = 2\mathcal{N} \left(\frac{d\ln
  C^{TE}_{\ell}}{d\ln p}\right)^2 \frac{r^2_{\ell}}{1 + r^2_{\ell}}.
  \label{eq:tandesignaltonoise}
\end{equation}
As before we have defined
$r_{\ell} \equiv C^{TE}_{\ell}/\sqrt{C^{TT}_{\ell}C^{EE}_{\ell}}$, which
is plotted in Figure~\ref{fig:corr}. For a scaling parameter we can simply
replace the derivative with unity.

For a general parameter $p$ there can be a different situation than
we saw for $TT$ and $EE$.  In the $TE$ power spectrum the
amount of correlation {\it and} (possibly surprisingly) how it changes under
the influence of $p$ is important.  To see this we can re-write the
square of the SNR of $TE$ in terms of $r$, $TT$, and
$EE$ as
\begin{align}
  \left(\frac{S^2}{N^2}\right)^{TE}_{\ell} = 2\mathcal{N}
  \frac{r^2_\ell}{1 + r^2_\ell}
  \left[\left(\frac{d\ln r_\ell}{d\ln p}\right)^2\right.
  &+ \frac{d\ln r_\ell}{d\ln p}\left(\frac{d\ln C^{TT}_\ell}{d\ln p} +
  \frac{d\ln C^{EE}_\ell}{d\ln p}\right)  \notag\\
  &\qquad\qquad+ \left.\frac{1}{4}\left(\frac{d\ln C^{TT}_\ell}{d\ln p}
  + \frac{d\ln C^{EE}_\ell}{d\ln p}\right)^2\right].
  \label{eq:tes2n}
\end{align}
We can then look at the two limiting cases $r_\ell \rightarrow \pm1, 0$:
\begin{align}
  \lim_{r_\ell\rightarrow \pm1}\left(\frac{S^2}{N^2}\right)^{TE}_{\ell}
  &= \frac{\mathcal{N}}{4}\left(\frac{d\ln C^{TT}_\ell}{d\ln p}
  + \frac{d\ln C^{EE}_\ell}{d\ln p}\right)^2;
  \label{eq:ter1} \\
  \lim_{r_\ell\rightarrow 0}\left(\frac{S^2}{N^2}\right)^{TE}_{\ell}
  &= 2\mathcal{N} \left(\frac{dr_\ell}{d\ln p}\right)^2.
  \label{eq:ter0}
\end{align}
The first thing to notice is that in the limit of full correlation the
information in $TE$ is directly given by the information content of $TT$ and
$EE$, as expected.  However, we also see that even in the case of a vanishing
$r_\ell$, there is still information to be obtained from measuring $TE$
(this is because the SNR does {\it not\/} vanish as $r_\ell\to0$).
This will depend on the behaviour of the parameter $p$ and specifically on
how $r_\ell$ varies as the parameter changes.
For a scaling parameter the
information content of $TE$ vanishes when $r_\ell$ vanishes, but
this is not true in general.

Details of the parameter ${\rm SNR}^2$ values per multipole are shown in
Figure~\ref{fig:SNRparams}, with the green curves being for the $TE$ case.
Once again, the $A_{\rm s}$ panel is helpful in here because we can see that
the dramatic drops in the green curve correspond to the points with
$r_\ell = 0$. Checking the green curve value of these locations at other
panels, such as $\theta_\ast$, reveals that one gains information by measuring
$TE$ even if $r_\ell = 0$ at a specific angular scale (e.g.,
see Figure~\ref{fig:SNRparams} just below $\ell = 1000$).

The $TE$ curves are usually lower than the $TT$ and $EE$ curves, simply
because the $T$- and $E$-modes are only partially correlated.  This is
manifested particularly strongly in the $\tau$ panel of
Figure~\ref{fig:SNRparams}, where it is clear that $TE$ is much less sensitive
to the reionization bump than $EE$.
However, we see that $TE$ can be higher in some multipole ranges for some
parameters, particularly for $\Omega_{\rm b} h^2$ and $\Omega_{\rm c} h^2$.
For example, $TE$ is more sensitive to
$\Omega_{\rm b} h^2$ than $TT$ in the range $670 \lesssim \ell \lesssim 1920$
and more sensitive than $EE$ in the range $810 \lesssim \ell \lesssim 1830$.
In fact Ref.~\cite{Galli14} already pointed out that $TE$ can constrain
$\Omega_{\rm c} h^2$ {\it better\/} than $TT$ by around $15\,\%$ -- this
would be hard to determine directly from Figure~\ref{fig:SNRparams}, since the
figure does not account for correlations among parameters (although this is
something we do consider in Section~\ref{sec:Params}).

\subsection{\textit{TT} and \textit{EE} power spectra, no correlation}
We would like to understand the basic way that polarization information
combines with temperature information.  So let us now consider the simple
(although hypothetical) situation in which we have
mapped out $E$-mode polarization, but when this
polarization is {\it uncorrelated\/} with temperature anisotropies.
Under these conditions
the covariance matrix of Eq.~(\ref{eq:cov}) takes on a simple $2\times2$ form.
In this case it is easy to verify that the information is exactly
doubled compared with the temperature-only case:
\begin{align}
  \left(\frac{S^2}{N^2}\right) &= \sum_{\ell}
  \left[ \left(\frac{S^2}{N^2}\right)^{TT}_{\ell}
  + \left(\frac{S^2}{N^2}\right)^{EE}_{\ell} \right].
  \label{eq:totalsignocorr}
\end{align}
Therefore for a scaling parameter the total SNR is
\begin{eqnarray}
\left( \frac{p}{\Delta p}\right)  &=&pF_{pp}^{1/2}  \notag \\
&\simeq&f_{\rm sky}^{1/2}\ell_{\max},
\end{eqnarray}
i.e., we obtain a factor of $\sqrt{2}$ improvement over the temperature-only
case.  This makes sense, because uncorrelated $E$-mode polarization is adding
precisely one additional piece of information for every pixel on
the sky, or equivalently, is adding an independent set of $a_{\ell m}^E$ modes
to the $a_{\ell m}^T$ modes.

\subsection{\textit{TT} and \textit{EE} correlated, but ignoring \textit{TE}}
The situation in the previous subsection is of course not realistic, because
in reality there is a $TE$ correlation in the CMB anisotropies, and hence there
are {\it three\/} distinct power spectra to determine, $C_\ell^{TT}$,
$C_\ell^{EE}$, and $C_\ell^{TE}$.  So how does this affect the total
information content?

To see how this works, let us first of all imagine that
although the temperature and polarization fields are correlated, we have
{\it not\/} measured this correlation (we are imagining an impractical scenario
here where the cosmologist has been careless and ignored $TE$).
We will again treat the simple case of a scaling parameter.
If we just use the $2\times 2$ matrix in
the upper left part of Eq.~(\ref{eq:cov}), we find
\begin{eqnarray}
 F_{pp} &=&\frac{2}{p^{2}}f_{\rm sky}\sum_{\ell =2}^{\ell_{\max}}
\left[ \left( \ell +\frac{1}{2}\right) \frac{1}{1+r_\ell^2}\right]\notag \\
 & \simeq &\frac{1}{p^{2}}f_{\rm sky}
 \left( \ell_{\max}^{2}{M}\right) .
\end{eqnarray}
Here we have defined
\begin{equation}
 {M}\equiv\frac{\sum_{\ell=2}^{\ell_{\max}}
 \left[\left(\ell+\frac{1}{2}\right)/(1+r_\ell^2)\right]}
 {\frac{1}{2}\ell_{\max}^{2}},
\label{eq:M2}
\end{equation}
giving the ratio between this case and the (previously considered) case where
$T$ and $E$ are uncorrelated.  Since $0<r^2_\ell<1$, it is clear that $1/2<{M}<1$
(at least for sufficiently high $\ell$).
The limit ${M}=1/2$ (equivalent to $r_\ell^2=1$) corresponds to a perfect
correlation between temperature and polarization and is therefore the same as
the temperature-only case.  On the other hand, ${M}=1$ (corresponding
to $r_\ell^2 = 0$), is when $T$ and $E$ are uncorrelated.

The signal-to-noise ratio for our hypothetical scaling parameter $p$ when
$C_\ell^{TE}$ is unmeasured is just
\begin{eqnarray}
 \left( \frac{p}{\Delta p}\right)  &=&pF_{pp}^{1/2}  \notag \\
 &\simeq&f_{\rm sky}^{1/2}\ell_{\max}{M}^{1/2}.
\end{eqnarray}

\begin{figure}[htb!]
\centering
\includegraphics[width=8.5cm]{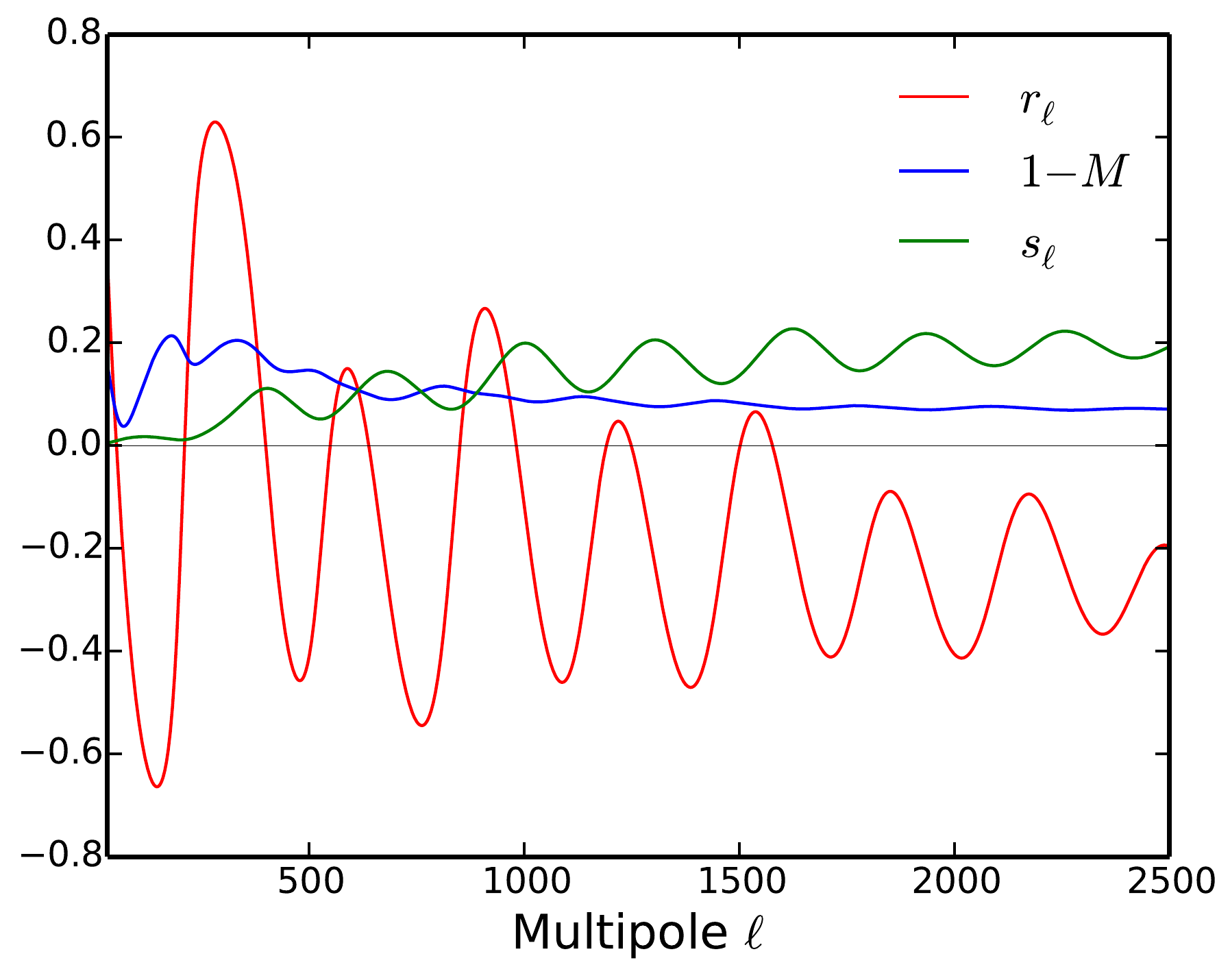}
\caption {Various quantities related to the correlations between the $T$ and
$E$-modes in the CMB sky, for best-fit 6-parameter $\Lambda$CDM model found by
{\it Planck\/} \cite{PlanckXIII}.  The red line shows the correlation
coefficient $r_\ell \equiv C_\ell^{TE}/(C_\ell^{TT}C_\ell^{EE})^{1/2}$, while
the green line shows the fractional polarization, defined as
$s_\ell \equiv (C_\ell^{EE}/C_\ell^{TT})^{1/2}$.  The blue line is the
quantity $M$ (defined in Eq.~\ref{eq:M2}), which is effectively the fraction
of the information lost by not observing $C_\ell^{TE}$ (along with
$C_\ell^{TT}$ and $C_\ell^{EE}$), plotted as a function of $\ell_{\rm max}$.}
\label{fig:corr}
\end{figure}

For the standard $\Lambda$CDM cosmology, we show the correlation coefficient
$r_\ell$ in Figure~\ref{fig:corr}, illustrating that for some multipoles the
magnitude of the correlation between $T$ and $E$ can be as high as 60\,\%.
We also plot $1-M$ as a function of $\ell_{\rm max}$; this effectively shows
the amount of information that would be lost by neglecting the
$TE$ cross-correlation power spectrum.  The plot shows that this can be as
much as 20\,\%, but for high $\ell_{\rm max}$ is a little under 10\,\%.

\subsection{Full \textit{TT}, \textit{EE}, and \textit{TE}}
\label{sec:fullfisherinformation}
Now let us consider the case when $C_\ell^{TT}$, $C_\ell^{EE}$, and
$C_\ell^{TE}$ are {\it all\/} measured (for a smiple scaling parameter first).
We then need to consider the full $3\times 3$ covariance matrix when
calculating the Fisher information and we recover
\begin{equation}
\left(\frac{p}{\Delta p}\right) \simeq f_{\rm sky}^{1/2}\ell_{\rm max},
\end{equation}
i.e., the same as in the situation where $T$ and $E$ are assumed
(unrealistically) to be uncorrelated.

This can be seen explicitly by inverting Eq.~(\ref{eq:cov}):
\begin{align}
 \mathbb{C}^{-1}_{\ell} &=
    \frac{2\mathcal{N}}{{|C|}^2}\times \left(
    \begin{array}{ccc}
      \frac{1}{2}{C^{EE}_{\ell}}^2 & \frac{1}{2}{C^{TE}_{\ell}}^2 &
-C^{EE}_{\ell}C^{TE}_{\ell} \\
      \frac{1}{2}{C^{TE}_{\ell}}^2  & \frac{1}{2}{C^{TT}_{\ell}}^2 &
-C^{TT}_{\ell}C^{TE}_{\ell} \\
      -C^{EE}_{\ell}C^{TE}_{\ell} & -C^{TT}_{\ell}C^{TE}_{\ell} &
C^{TT}_{\ell}C^{EE}_{\ell} + {C^{TE}_{\ell}}^2
    \end{array} \right) \nonumber\\
    &= \frac{\mathcal{N}}{{|C|}^2}C^{TT}_{\ell}C^{EE}_{\ell}\left(
    \begin{array}{ccc}
      s^2_\ell & r^2_{\ell} & -2r_{\ell}s_\ell \\
      r^2_{\ell} & s^{-2}_\ell & -2r_{\ell}/s_\ell \\
      -2r_{\ell}s_\ell & -2r_{\ell}/s_\ell & 2\left(1 + r^2_{\ell}\right)
    \end{array} \right),
\label{eq:invcov}
\end{align}
where the quantity $s_\ell$ has been introduced as the ratio of the
polarization to temperature anisotropy and we also define $\mathcal{N}$ and $|C|$ to
simplify the expression above, specifically with
\begin{eqnarray}
  \mathcal{N} &\equiv& \frac{(2\ell + 1)f_{\rm sky}}{2}, \\
  |C| &\equiv& C^{TT}_{\ell}C^{EE}_{\ell} - {C^{TE}_{\ell}}^2 \\
    &=& C^{TT}_{\ell}C^{EE}_{\ell}\left(1 - r_{\ell}^2\right), \\
  s_\ell^2 &\equiv& \frac{C^{EE}_{\ell}}{C^{TT}_{\ell}}.
\end{eqnarray}
The quantity $s_\ell$ is also plotted in Figure~\ref{fig:corr}, showing that
polarization anisotropies are only a few percent of temperature anisotropies
at large angular scales, and asymptote to a value close to 20\,\% at higher
multipoles.

Following the algebra introduced above, the Fisher matrix becomes
\begin{align}
  F_{pp} &= \frac{1}{p^2} \sum_{\ell} \left(
    \begin{array}{c}
      C^{TT}_{\ell} \\
      C^{EE}_{\ell} \\
      C^{TE}_{\ell}
    \end{array} \right)^{\sf T} \mathbb{C}^{-1}_{\ell} \left(
    \begin{array}{c}
      C^{TT}_{\ell} \\
      C^{EE}_{\ell} \\
      C^{TE}_{\ell}
    \end{array} \right) \\
    &= \frac{1}{p^2} \sum_{\ell} C^{TT}_{\ell}C^{EE}_{\ell} \left(
    \begin{array}{c}
      s^{-1}_\ell \\
      s_\ell \\
      r_{\ell}
    \end{array} \right)^{\sf T} \mathbb{C}^{-1}_{\ell} \left(
    \begin{array}{c}
      s^{-1}_\ell \\
      s_\ell \\
      r_{\ell}
    \end{array} \right) \\
    &= \frac{1}{p^2} \sum_{\ell} \frac{\mathcal{N}}{\left(1 - r^2_{\ell}\right)^2} \left(
    \begin{array}{c}
      s^{-1}_\ell \\
      s_\ell \\
      r_{\ell}
    \end{array} \right)^{\sf T} \left(
    \begin{array}{c}
      s_\ell(1 - r^2_{\ell}) \\
      s^{-1}_\ell(1 - r^2_{\ell}) \\
      -2r_{\ell}(1 - r^2_{\ell})
    \end{array} \right) \\
    &= \frac{2}{p^2} \sum_{\ell} \mathcal{N} \frac{\left(1 - r^2_{\ell}\right)^2}{\left(1
  - r^2_{\ell}\right)^2} \\
    &= \frac{2}{p^2}f_{\rm sky}\sum_{\ell} \left(\ell + \frac{1}{2}\right),
\label{eq:finalFisher}
\end{align}
which is similar to the result of the previous section, with $M = 1$.
Therefore, measurements to some $\ell_{\rm max}$ of the full $TT$, $EE$, and
$TE$ spectrum have {\it exactly\/} twice as much information as a
temperature-only (or polarization-only) experiment.
In appendix~\ref{app:decorrelation} we give a conceptually simpler
derivation of this same
result, by transforming to fields that are uncorrelated by construction.

What about more complicated parameters?
For power spectra with arbitrary dependence on a single parameter $p$ the
total squared SNR is then
\begin{align}
  \left(\frac{S}{N}\right)^2 &= \sum_{\ell} \frac{1}{\left(1-r^2_\ell\right)^2}
  \left\{\left(\frac{S^2}{N^2}\right)^{TT}_{\ell} +
  \left(\frac{S^2}{N^2}\right)^{EE}_{\ell}
   + 2r^2_\ell\left(\frac{S}{N}\right)^{TT}_{\ell}
  \left(\frac{S}{N}\right)^{EE}_{\ell}\right. \notag\\
  &\quad\left. -
   4r_\ell\sqrt{2+2r^2_\ell}\left[\left(\frac{S}{N}\right)^{TT}_{\ell}
  + \left(\frac{S}{N}\right)^{EE}_{\ell}\right]
  \left(\frac{S}{N}\right)^{TE}_{\ell}
  + 4\left(1+r^2_\ell\right)^2
  \left(\frac{S^2}{N^2}\right)^{TE}_{\ell}\right\}.
  \label{eq:singleparameter}
\end{align}
In appendix~\ref{app:decorrelation} we present an interpretation of the
contributions to the total SNR, which come from temperature (uncorrelated with
polarization), polarization (uncorrelated with temperature), and the
correlation itself, $\theta_{\ell}$ (defined in Eq.~\ref{eq:theta}).
Equation~\ref{eq:singleparameter},
however allows us to consider the situation at a scale $\ell$ for which
$r_\ell = 0$.  In this case we obtain
\begin{align}
  \left(\frac{S}{N}\right)^2_{\ell} &= \left(\frac{S^2}{N^2}\right)^{TT}_{\ell}
  + \left(\frac{S^2}{N^2}\right)^{EE}_{\ell} + 4
  \left(\frac{S^2}{N^2}\right)^{TE}_{\ell},
  \label{eq:singleparameternocorrelation}
\end{align}
where the final term is given by 4 times Eq.~(\ref{eq:ter0}). We see here
(somewhat surprisingly) that the $TE$ correlation will contribute information
even when $r_{\ell} = 0$, provided that $d r_{\ell}/dp \neq 0$.

While Eqs.~(\ref{eq:singleparameter}) and (\ref{eq:decorrelatedston}) are
complete descriptions of the contributions to the information content of a
single arbitrary parameter, they do not contain the details of how multiple
parameters
are constrained or how parameter correlations are involved, which we consider
in Section~\ref{sec:Params}.

\subsection{Total SNR for a single parameter}
To complete this section, let us be more explicit, and give
a quantitative example.  For a cosmic-variance-limited experiment up to
$\ell_{\rm max}=3000$ and covering the entire sky, the total signal-to-noise
ratio in a single scaling parameter is $S/N \simeq 2100$.
And if we add $E$-mode polarization information, also cosmic-variance-limited
to the same $\ell$, we obtain $S/N \simeq 3000$.

An experiment that makes ideal measurements of $T$
and $E$ out to some $\ell_{\rm max}$ has precisely twice as much information
(i.e., constraining power for a scaling parameter) as a $T$-only
experiment, provided that all of $TT$, $EE$, {\it and\/} $TE$ are measured.
A parameter with more complicated $C_\ell$ dependence, like $\theta_\ast$ for
example, will have more information from $EE$ due
to its stronger contrast between peaks and troughs. Such a parameter will also
have part of its constraint coming from how the correlation itself changes (an
effect that cannot be seen with a simple scaling parameter).

In terms of constraints on specific parameters, we know that the situation is
more complicated still.  For example, polarization data
are important for breaking particular degeneracies
\citep{Zaldarriaga97,Seljak97} (especially for
determining the reionization optical depth, $\tau$), and so polarization may
constrain some parameters much better than expected for simply twice as much
information.  We have already seen this presented in Figure~\ref{fig:SNRparams}
In Section~\ref{sec:Params} we will focus on several $\Lambda$CDM
parameters and how their correlations affect parameter constraints.

\section{Additional CMB information}
\label{sec:other}
Before investigating parameter dependence in detail, it is worth noting that
some other information that could come from the CMB, in addition to the
three power spectra we have been considering.

If one could measure $BB$ to the same $\ell_{\rm max}$ then instead of the
full $TT$ + $TE$ + $EE$ measurement giving twice as many modes as come from
$TT$ alone, we would now have {\it 3 times\/} as many modes.  In practice we
expect primordial $B$-modes to be quite weak, and it is extremely unlikely
that we could measure this power spectrum beyond the first few hundred
multipoles (e.g., Ref.~\cite{simard}).  Hence $B$-modes are never going to add
substantially to the mode count.  On the other hand, {\it any\/} measurement of
primordial $B$-modes would provide a direct constraint on the tensor-to-scalar
ratio that would be better than the indirect constraints from other power
spectra.  Hence (as is well known) the constraints on this additional parameter
are dramatically improved through better $B$-mode experiments.

We have been assuming that the CMB sky contains Gaussian perturbations, but
we know that this cannot be exactly true.  Certainly there is hope that we
may one day detect non-Gaussianity from higher-order correlations in
the CMB (e.g., Ref.~\cite{PlanckXVII}), and here polarization offers the
promise of pushing the uncertainties down.  However, we do not expect such a
signal to give a very high SNR (at least compared to the power spectra),
since the CMB is clearly very close to Gaussian.

One exception to this is that the 4-point function of the CMB sky contains
correlations from the effects of gravitational lensing.  This signature
allows us to estimate an additional power spectrum, $C_L^{\phi\phi}$,
which has already been done to ${\rm SNR}\simeq40$ by {\it Planck\/}
\cite{PlanckXV}.  If we could measure this power spectrum to the same
maximum multipole as for the temperature and polarization power spectra, then
it would add the same number of modes again.  However, things are not so
simple, because the index $L$ for lensing comes from the coupling of
modes at different scales, and hence making a noise-free temperature map out to
$\ell_{\rm max}$ will not give a cosmic variance limited measurement of
$C_L^{\phi\phi}$ to $L=\ell_{\rm max}$.  But putting that aside, lensing
will add effectively as many modes as temperature and polarization.  Hence
there is in principle $\sqrt{3}$ times as much information contained in a
full CMB mapping experiment (plus some additional information from $B$-modes)
as there is in a measurement purely of $C_\ell^{TT}$.
On the other hand, in terms of parameter constraints, the lensing power
spectrum has little dependence on cosmological parameters other than
amplitude \cite{Smith06,Challinor07}.

Since $\phi$ is a Gaussian random field, its power spectrum
has the same statistical properties as the $T$ or $E$ maps. Therefore,
adding lensing to the covariance matrix is similar to adding $EE$ or
$TE$ to $TT$.
One could also consider adding the information coming from lensed $B$-modes
-- but then one would have to account for the fact that these modes are
{\it not\/} Gaussian (since they come from a convolution of $E$-modes with
lensing $\phi$-modes \cite{Antony_lensing}).
However, the addition of lensed $B$-modes provides little to no improvement on
parameters if $E$-modes and lensing $\phi$-modes are already accounted for. The
reason is an intuitive one: the lensed $B$-modes come directly from
{\it un\/}lensed $E$-modes and $\phi$-modes. Thus adding $B$-modes simply
double counts some combination of the $E$- and $\phi$-modes (albeit at
different scales). Thus adding the lensed $B$-modes only helps in the
noise-dominated case.
In Section~\ref{sec:neutrino} we consider including lensing $\phi$-modes to our
data vector (Eq.~\ref{eq:Fisher}); for the reasons stated above we do not
simultaneously consider lensed $B$-modes.

\section{CMB parameter information}
\label{sec:Params}
\subsection{Relationship with overall SNR}
We now discuss the connection between the information in the power spectra
and the constraints on the 6-parameters of the standard $\Lambda$CDM model.
One might expect the total SNR in $\ell$-space to be very crudely
of order the SNR in the parameter space; however, in detail we do not expect
them to be the same.  This is due to the degeneracies
between parameters and the sensitivity of subsets of the data to changes in
specific parameters, as well as the important fact that the power spectra do
not depend linearly on the parameters, as was discussed already
(see Figure~\ref{fig:SNRparams}).

\begin{figure}[htb!]
\centering
\includegraphics[width=8.5cm]{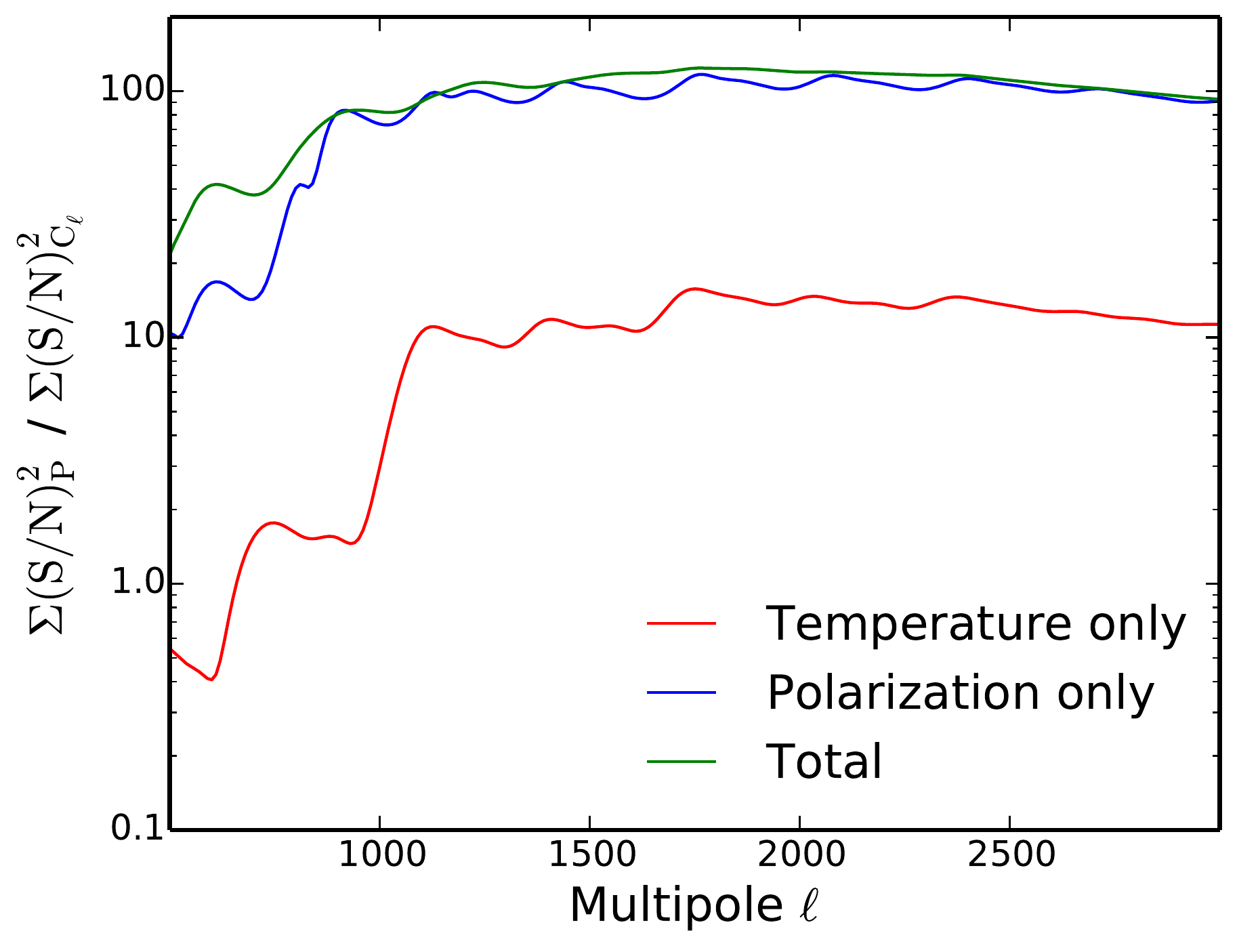}
\includegraphics[width=8.5cm]{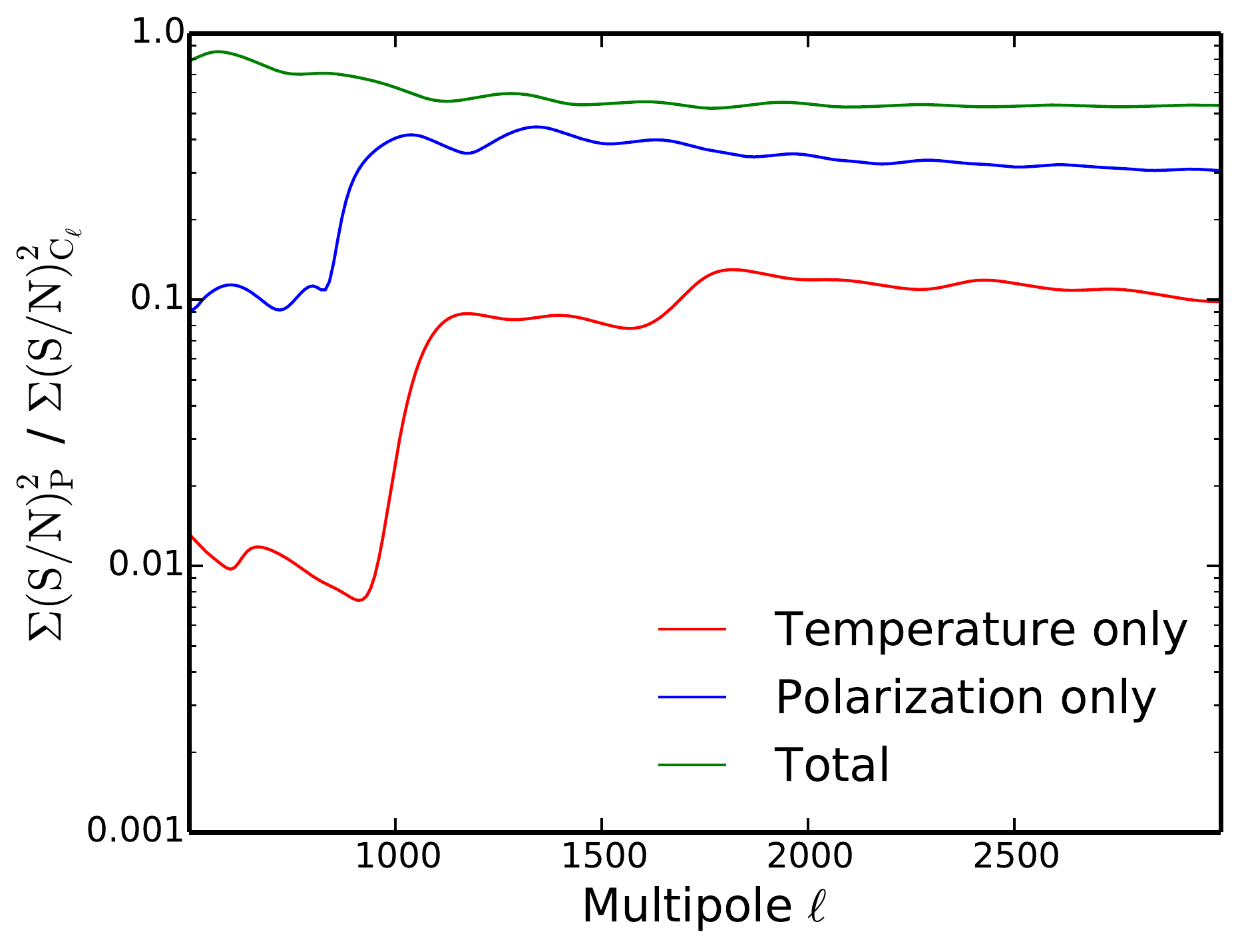}
\caption {{\it Top}: ratio of the total ${\rm SNR}^2$
in parameters to that in the power spectra for a cosmic-variance-limited
experiment.  Note that this quantity is independent of the sky coverage.
The red line shows the results of a temperature-only experiment,
the blue line is for a polarization-only experiment, and the green
line shows the results for both temperature and polarization combined.
The parameter set here consists of
$\{\Omega_{\rm b}h^2,\Omega_{\rm c}h^2,\theta_\ast,\tau,A_{\rm s},n_{\rm s} \}$.
{\it Bottom}: same as the top panel, except that $H_0$ has been used
as a parameter in place of $\theta_\ast$, i.e.,
the parameter set here consists of
$\{\Omega_{\rm b}h^2,\Omega_{\rm c}h^2,h,\tau,A_{\rm s},n_{\rm s} \}$.
}
\label{fig:SN}
\end{figure}

One can think of the $N_{\rm p}$-dimensional parameter space as a
data compression scheme.  This compression reduces the $\ell$ measured power
spectrum values and their uncertainties into $N_{\rm p}$ numbers and their
corresponding uncertainties.  The power spectra depend linearly on the
parameter $A_{\rm s}$ (apart from small lensing effects), and hence if that was
the only parameter, then its
SNR would be the same as the SNR for the power spectra, and hence would be the
same as the mode counting exercise discussed in the section~\ref{sec:Fisher}.
However, other parameters are not ``linear'' in this sense, and hence can be
constrained better or worse than seen for the simple case of $A_{\rm s}$.

As a simple example, if someone wanted to treat $A_{\rm s}^2$ as a parameter
instead of $A_{\rm s}$, then the SNR would be better by a factor of 2.
A more dramatic change will come from considering $\ln A_{\rm s}$ rather than
$A_{\rm s}$.  And in general the standard parameters affect the power spectra
in ways that are fairly different from those of a scaling parameter.
The most non-linear of the parameters is $\theta_\ast$, as was already
discussed, since fairly small changes in $\theta_\ast$ can result in large
changes to the power spectra, because of the relative sharpness of the
adiabatic peaks and troughs.  The best way
to understand the constraints on parameters in general is to use
the Fisher matrix to investigate what happens for the standard 6-parameter
cosmology.  In general, however, we should not be surprised to find that the
SNR values for some cosmological parameters only differ by factors of order
unity from those of a linearly scaling parameter.

A comparison of the total SNR in parameters with the total SNR in the power
spectra is shown in Figure~\ref{fig:SN}. Here the
error bars for each parameter are derived from a Fisher matrix calculation,
assuming a noise-free experiment with sample variance only, with sky coverage
of 50\,\% (picked to approximately match that of {\it Planck\/}).
The fiducial model used for the Fisher matrix calculation is $\Lambda$CDM
with parameters equal to the best-fit values of the \textit{Planck}-2015
``TT+lowTEB+lensing'' combination \cite{PlanckXIII}.
The total SNR in parameter $p_i$ is
\begin{equation}
\left(\frac{S}{N}\right)^2_i
 = \sum_{\ell} \, {p^2_i \over (F_{\ell}^{-1})_{ii}},
\end{equation}
and Eq.~\ref{eq:SN_Cl} (including an $f_{\rm sky} = 0.5$ factor) is used to
calculate the total SNR in power spectra.

Figure~\ref{fig:SN} (top panel) shows that the total ${\rm SNR}^2$ in
parameters can be a factor of as much as 100 larger than the total
${\rm SNR}^2$ in the power
spectra.  However, this number is highly dependent on the set of parameters
that are chosen and can vary dramatically, e.g., if $\theta_\ast$ is
replaced by
$H_0$, as plotted in the bottom panel of Figure~\ref{fig:SN}. The ratio of the
total SNR in parameters to the SNR in the power spectra (temperature and
polarization) is close to unity when $H_0$ is substituted for $\theta_\ast$.
This suggests that the combination of parameters in the bottom panel of
Figure~\ref{fig:SN} is more ``linearly'' related to the power spectra
than the set in the top panel.  The behaviour seen in this figure explains
the observation made in Section~\ref{sec:Intro} that the SNR on $\theta_\ast$
from {\it Planck\/} exceeds the total SNR in the $TT$ power spectrum.

One can learn more about parameter constraints by looking more closely at
Figure~\ref{fig:SN}.  Firstly, the relatively poor constraints on the
6-parameter set at low multipoles is a result of parameter degeneracies, which
are not broken until higher multipole data are included.
Secondly, we see that adding polarization data makes a
substantial improvement to the overall constraint on parameters.  The more
dramatic improvement in the total constraint (green lines in the figure)
at low multipoles is a result of the well-known ability of CMB polarization
data to break the $A_{\rm s}e^{-2\tau}$ degeneracy.  Thirdly, the polarization
data on their own are more constraining (for the same $\ell_{\rm max}$) than
the temperature data; this arises essentially because the polarization power
spectra are ``sharper'' than for $C_\ell^{TT}$, as has been stressed in
other studies (e.g., Ref.~\cite{Galli14}).  We also see structure in
these SNR curves that clearly reflects the shape of the $C_\ell$s; we shall
discuss this further in the next section.

\subsection{Parameter constraints from power spectra}
Although the SNR in the power spectra gets effectively shared out among the
parameters, as we have seen this is only very crudely correct, and in
practice the detailed constraints on the parameters will change based on many
factors.  In particular the parameter constraints will depend on
which power spectra are used, which
multipole range is measured, and what set of parameters was chosen in the
first place.  We need to appreciate that
some information is special for particular parameters, e.g., large-angle
polarization for $\tau$; thus we should focus on the $\ell$-range that is
important for each parameter.  We now describe this connection more
comprehensively by showing some
examples and comparing with the current uncertainties.  It is important
to realize that we are not intending here to make forecasts for specific
experiments (with particular assumptions about beamsize, noise, foreground
contamination, etc.), since this has been done before.  Instead we are
asking the more general question of how good the parameter constraints could
{\it ever\/} become, by comparing the ideal values with current constraints
from {\it Planck}.

\begin{figure*}[htb!]
\centering
\includegraphics[width=\textwidth]{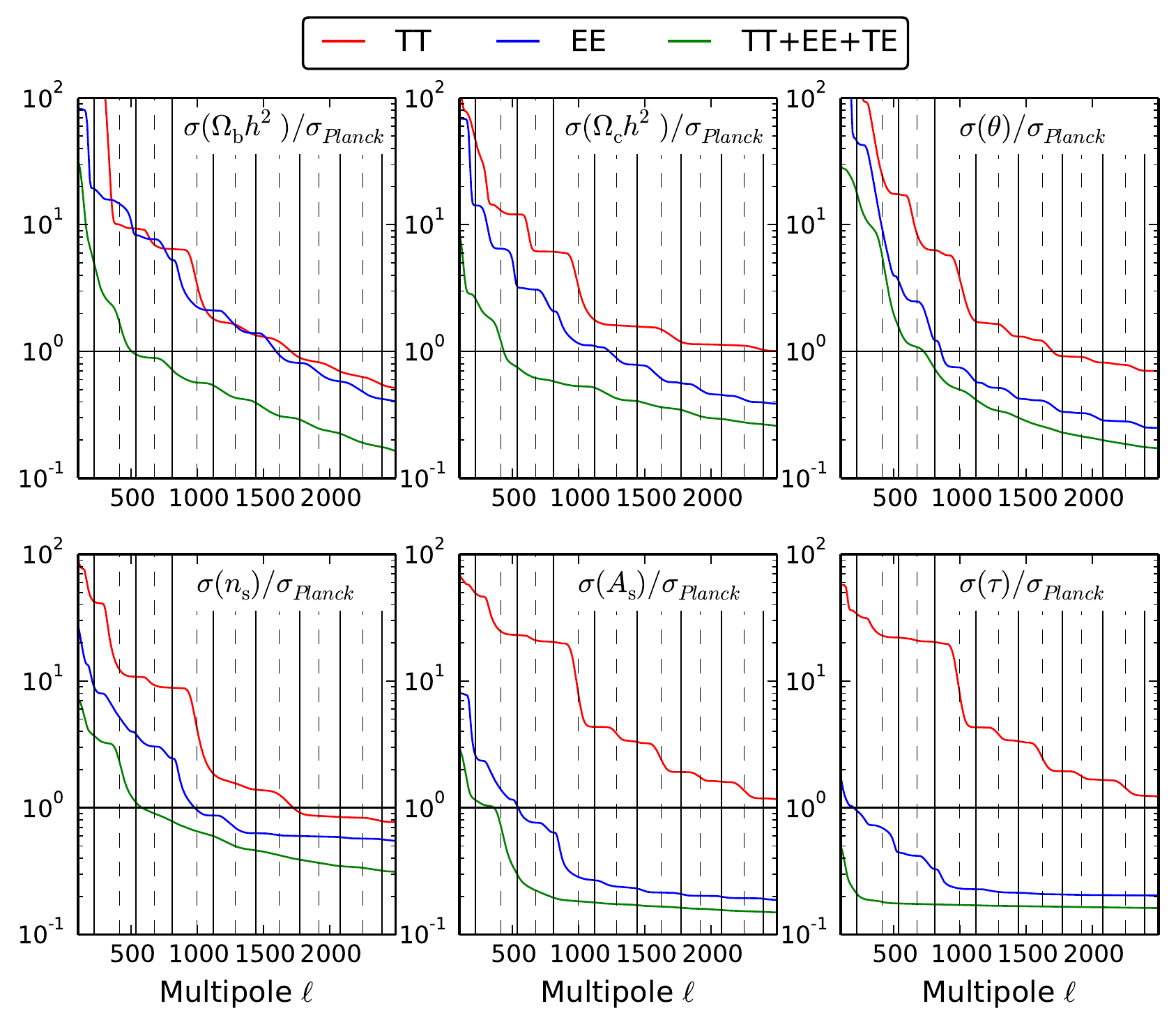}
\caption {Cumulative Fisher forecast calculations for the six parameters of
the standard $\Lambda$CDM model for a noise-free experiment with
$f_{\rm sky}=0.5$.  The red line is for a temperature-only
experiment, the blue line is for pure polarization, and the green line is for
the full power spectra ($TT$, $TE$, and $EE$). The error bars for each
parameter are compared with the \textit{Planck}-2015 ``TT+lowTEB+lensing''
68\,\% confidence limits at different values of maximum multipole. The solid
vertical lines indicate the
temperature peak positions, while the dashed lines are for temperature
troughs. Note that the peaks (troughs) in the $TT$ power spectrum are in
almost the same positions as the troughs (peaks) in the $EE$ spectrum.}
\label{fig:6params_lensed}
\end{figure*}

Figure~\ref{fig:6params_lensed} shows the results of a Fisher calculation for
a cosmic-variance-limited, CMB-only experiment, where we have picked a sky
coverage of 50\,\%, which approximately matches the effective area used
by {\it Planck\/} for the main parameter constraints (although this is
frequency dependent, see Ref.~\cite{PlanckXIII}).  Note that
the errors would just scale as $f_{\rm sky}^{-1/2}$
for other values.  The $x$-axis here indicates the maximum $\ell$ used in the
calculation. The red line is for a temperature-only experiment, while the blue
and green lines are for polarization-only and the full set of three spectra,
respectively. The error bars for each
parameter are compared with the \textit{Planck}-2015 ``TT+lowTEB+lensing''
68\,\% confidence limits.  The vertical solid and dashed lines show the
temperature peak and trough positions, respectively, which are almost the same
as the $E$-polarization troughs and peaks \cite{PlanckXI}.

The mode-counting argument (of Section~\ref{sec:CMB})
tells us that every $a_{\ell m}$ has equal
weight for contributing to the SNR of the power spectrum, but, as we have seen,
this is not true for the parameter SNR, since the $C_\ell$s are not equal in
delivering constraints on individual parameters.  For example, while
the multipoles around the third $TT$ trough, $\ell\simeq 900$--1100, are
particularly useful for reducing the uncertainty on $\Omega_{\rm c} h^2$,
the information gain from the $\ell\simeq 1300$--1650 range (from
the fourth $TT$ peak to the fifth trough) is almost negligible.

\begin{figure*}[htb!]
\centering
\includegraphics[width=\textwidth]{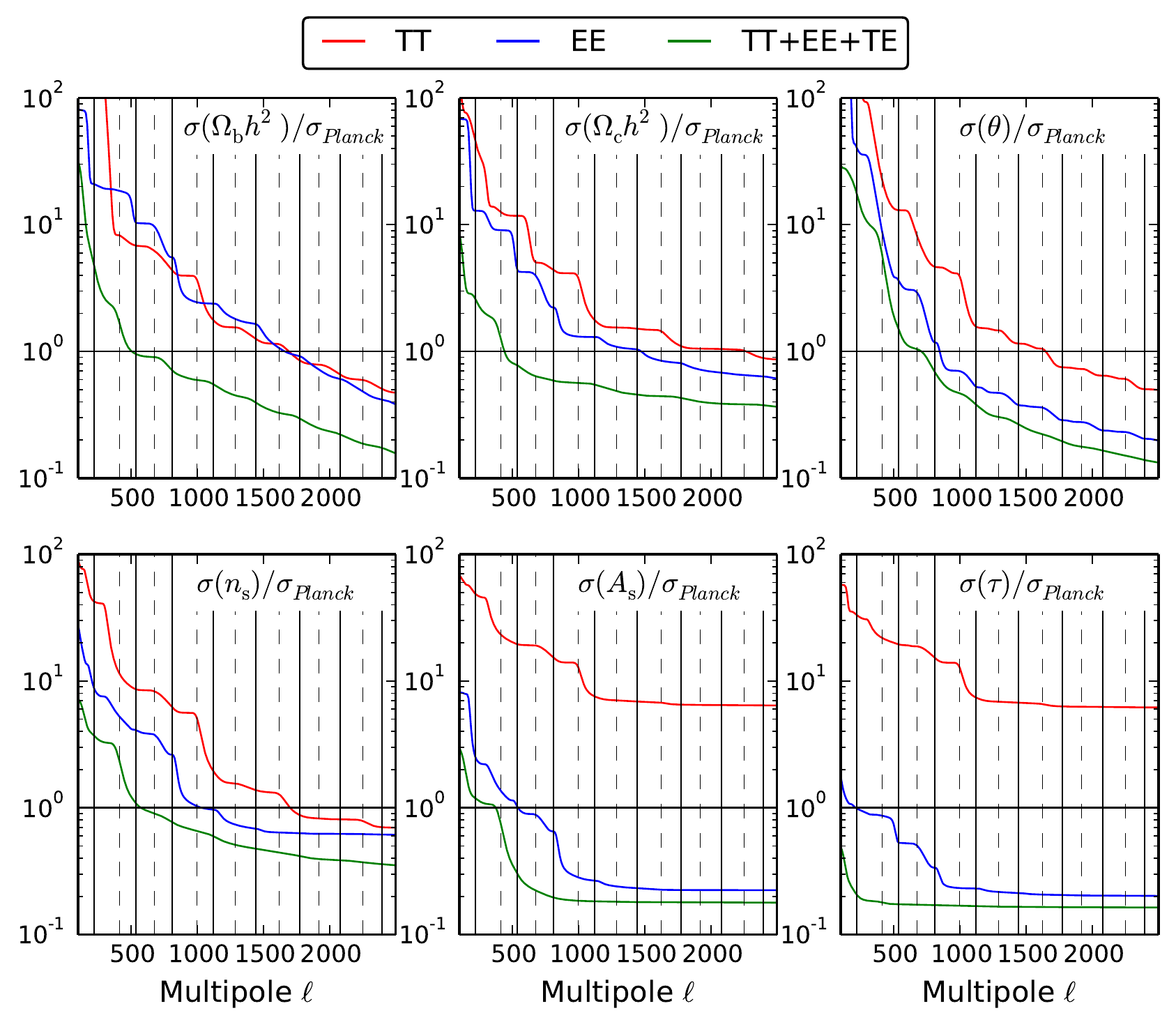}
\caption {Cumulative Fisher forecast calculations for the six parameters of the
standard $\Lambda$CDM model for a noise-free experiment with
$f_{\rm sky}=0.5$, using \textit{un\/}lensed spectra.
The red line is for temperature only, the blue line for polarization only,
and the green line for the full $TT$, $TE$, and $EE$ spectra. The $y$-axis here
is the ratio of the uncertainty on each parameter compared with the
\textit{Planck}-2015 ``TT+lowTEB+lensing'' 68\,\% confidence limits,
at different maximum multipoles. The vertical lines are the positions of
peaks (solid lines) and troughs (dashed lines) in the $TT$ power spectrum
(and opposite for the the $EE$ spectrum).}
\label{fig:6params_NotLensed}
\end{figure*}

A careful examination of Figure~\ref{fig:6params_lensed} shows that in general
the $TT$ {\it troughs\/} are more important than the peaks for reducing the
error bars (particularly clear when focusing on $A_{\rm s}$ and $\tau$,
for example). We also see (in the green lines) that when we add temperature to
polarization the curves are much smoother. This is
because the effects coming from the troughs and peaks of $TT$ more or less
cancel with the effects coming from the troughs and peaks of $EE$.

But there is still the issue to explain of why the $TT$ troughs, which are
obviously {\it lower\/} than the peaks, and therefore (one might expect)
carry less information, should give stronger constraints.
By comparing Figure~\ref{fig:6params_lensed} with
Figure~\ref{fig:6params_NotLensed} -- which shows the same predictions but with
lensing effects turned off -- one can see that the reason for the importance of
the troughs is the effect of lensing on the $TT$ spectrum.
The explanation is that since lensing smooths the peaks and troughs, while
preserving total power, then the {\it relative\/} change from lensing is
larger around the troughs than around the peaks, and hence the troughs can give
better constraints on parameters.

Comparing Figure~\ref{fig:6params_lensed} with
Figure~\ref{fig:6params_NotLensed} also shows that while the lensing of the
power spectra is useful for breaking the $A_{\rm s}$--$\tau$ degeneracy, it
makes the constraints on the set $\{\theta_\ast,\Omega_{\rm c}h^2,
\Omega_{\rm b}h^2 \}$ {\it weaker}. This is because the peaks and troughs are
``sharper'' in the unlensed spectra, and are therefore a better source of
information for constraining $\theta_\ast$.  The uncertainties on the matter
densities are also improved as a result of the better constraints on
$\theta_\ast$ and the correlations among the parameters.

What we have highlighted here is the simple observation that constraints on
cosmological parameters come from two basic factors: the first is having power
spectra that are sensitive to changes in parameters (see
Eq.~\ref{eq:singleparameter} and Figure~\ref{fig:SNRparams}); the second is
the ability for power spectra to break degeneracies between parameters. In the
following subsection we will consider the effect of adding lensing
$\phi$-modes, which
generally have weak dependence on cosmological parameters. Nevertheless,
these modes can break parameter degeneracies, which is crucial for going
beyond the 6-parameter model.

\subsection{Extended parameters -- neutrino mass}
\label{sec:neutrino}
\begin{figure}[htb!]
\centering
\includegraphics[width=8.5cm]{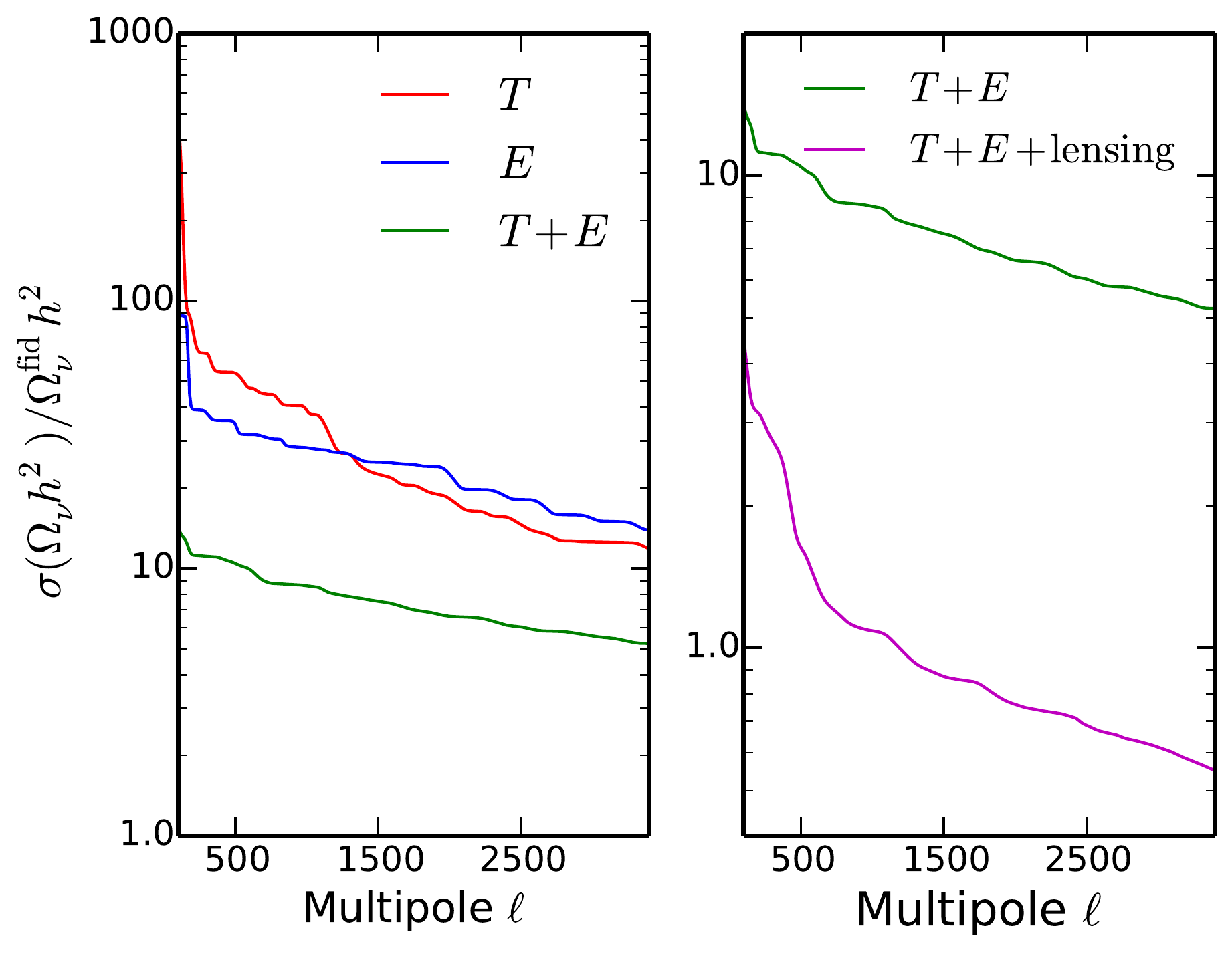}
\caption {Fisher calculation results for the total neutrino mass using CMB
data. The left panel compares the results of $TT$, $EE$, and the total
($TT+TE+EE$).  The right panel compares the results for $TT+TE+EE$
with the case when CMB lensing (including cross-power spectra) is also added.
It is assumed here that the CMB lensing spectra can be measured up to the
sample variance limit for $f_{\rm sky}=0.5$.  The fiducial neutrino density
used here is $\Omega_{\nu} h^2=0.00064$, which
corresponds to 60\,meV in mass.}
\label{fig:nu_mass}
\end{figure}

We will now choose a specific example to illustrate how we can think about the
relationship between information and parameter constraints.
Future CMB observations will target extensions to the 6-parameter
$\Lambda$CDM model.  In particular, the detection of the sum of the masses of
the neutrino species seems like a realistic (although challenging) possibility
in the near future \cite{LPPP06,Pan15,Allison15}, since the current upper
limit on $m_{\nu}^{\rm tot}$ \cite{PlanckXIII} is only a factor of a few
higher than the $0.06\,$eV limit imposed by direct measurements of
mass differences (e.g., Ref.~\cite{Feldman13}).  However, the left panel of
Figure~\ref{fig:nu_mass} shows that such a measurement is not feasible with
CMB temperature or polarization maps alone, because there
is not enough information in the power spectra out to $\ell\simeq3000$.

Besides the temperature and polarization fluctuations, one can also map
the fluctuations of the gravitational potential or, equivalently, the lensing
deflection angle.  Inclusion of the lensing power spectrum in the Fisher
formalism is discussed in Ref.~\cite{fisher_lensing}, assuming Gaussianity
and ignoring correlations between
different multipoles. Making similar assumptions, we find that
CMB lensing can improve the constraints on the total neutrino mass
dramatically, as is shown in the right panel of
Figure~\ref{fig:nu_mass}.  Here we have assumed
that the CMB lensing power spectrum, $C_\ell^{\phi\phi}$, and
its correlations with temperature and $E$-mode polarization can be measured
with sample-variance accuracy over $f_{\rm sky} = 0.5$ of the entire sky
(see e.g., Ref.~\cite{fisher_lensing}).
Given these assumptions, the full set of CMB power spectra reaches the
fiducial sensitivity (corresponding to $60\,$meV in mass) at
$\ell\simeq1200$,\footnote{Here we are using the same multipole symbol,
$\ell$, for lensing and temperature or polarization.}
and can ultimately measure the mass at about the 2$\,\sigma$ level.
However, there is a very strong degeneracy between the neutrino mass
and the dark matter density (with a correlation coefficient
$\simeq 0.95$ at $\ell_{\rm max}=2500$), as
well as a fairly strong degeneracy between neutrino mass and $n_{\rm s}$
(with a correlation coefficient of $\simeq 0.73$).
Hence, any additional independent measurement of these parameters -- e.g.,
via galaxy weak lensing, baryon acoustic oscillations, or redshift space
distortions -- that can break these degeneracies will lead to substantial
improvement.

\section{Discussion}
\label{sec:Disc}
In terms of mode-counting, the information contained in the CMB anisotropies
is clearly finite, because it is limited by cosmic variance and the fact that
the power spectra damp at the highest multipoles.
For temperature information alone the total SNR in the power spectrum is
$\sim \sqrt{f_{\rm sky}/2}\,\ell_{\rm max}$.  {\it Planck\/} has measured most of
what is available out to $\ell\simeq2000$, with ACT and SPT continuing that
out to higher $\ell$, but over relatively small $f_{\rm sky}$, and with
foregrounds making it seem unrealistic to push beyond $\ell\simeq3000$, say.
This means that although we can continue to measure our CMB sky to ever more
sensitive levels, we have already reached a point where the bulk of the
useful information has already been extracted from the temperature
anisotropies.

However, the situation is different for polarization
information, since the foregrounds (from galaxies and clusters of galaxies)
are very weakly polarized, and hence there is hope that we should be able to
measure the primary polarization anisotropies to much higher multipoles,
$\ell\simeq5000$ and perhaps considerably higher \cite{Benson14,Naess14}.
This means that there is at least an order of magnitude more polarization
information to extract from the CMB sky.  Moreover, as we have seen,
the polarization data can place tighter constraints on parameters in general,
and on some parameters in particular.  The $BB$ and $\phi\phi$ power spectra
add additional information, but in practice this is probably a small fraction
of what is available from the $T$- and $E$-modes.

Despite the dramatic improvement still expected from CMB polarization,
the maximal SNR in the CMB power spectra
is $\lesssim10^4$ from primary CMB anisotropies.  To improve cosmological
parameter constraints we therefore need to go to 3-dimensional surveys
(such as high-$z$ 21-cm fluctuations) where there are considerably more
modes \cite{MaScott}.
As an illustration of what this CMB limitation means, let us consider
the determination of the curvature of space, $\Omega_K$.  The
cosmic variance limit is at the $10^{-5}$ level, since this is the
amplitude of the curvature perturbation on the Hubble
scale~\cite{Waterhouse08}.  The current uncertainty from CMB data is at
the $10^{-2}$ level, and given the above argument, we expect this to only
decrease by about another order of magnitude.  Hence, assuming that the
Universe is sufficiently close to being spatially flat, then we
will never be able to determine whether it is flat or curved using CMB data
alone -- there is simply not enough information for us to achieve the
required SNR level on $\Omega_K$ from CMB anisotropies.
Ambitious future experiments may probe $\gtsim10^{12}$ 3D modes, which would
in fact allow us to reach below the cosmic-variance limit for the
measurement of $\Omega_K$ within our observable volume.

The discussion here has focused on the Gaussian primary anisotropies,
supplemented by CMB lensing.  However, we should acknowledge that there is
also some cosmological information content in the
secondary anisotropies (i.e., cosmic IR background, integrated
Sachs-Wolfe effect, Sunyaev-Zeldovich effects, etc.).
These effects certainly enable further cosmological information to be measured,
not just on the last-scattering surface, but also at other epochs along the
light cone.  Nevertheless the additional information seems limited in its
scope for constraining background parameters, because either there is only
a modest amount of information available at all (like in the ISW effect),
or the additional information is still effectively on a 2D surface.
The only way to obtain a dramatic improvement in constraining power will be
to pursue methods that are fundamentally 3D.

\section{Conclusions}
\label{sec:Concl}
We have taken a pedagogical approach to investigating the information content
in CMB anisotropies, in the sense of constraining the cosmological model.
It is clear that for temperature anisotropies, we have already mined a
substantial part of what is available, and we are effectively
running out of information.  However, for CMB polarization we still have a way
to go, and there may be an order of magnitude more constraining power still
to extract from the CMB sky.

The CMB $TT$ power spectrum has an SNR of approximately
$\sqrt{f_{\rm sky}/2}\,\ell_{\rm max}$, which can be thought of as a simple
mode-counting calculation.  The SNR on a scaling parameter (like $A_{\rm s}$)
is the same, while some parameters (such as $\theta_\ast$) have a dependence
which is ``non-linear,'' hence allowing them to be constrained more tightly
than the total SNR for the power spectrum as a whole.

We have shown that the mapping from information about the CMB power spectra
to information about the cosmological parameters has two ingredients, namely
the sensitivity of the power spectra to parameters
(Eq.~\ref{eq:singleparameter} and Figure~\ref{fig:SNRparams}), and the ability
to break parameter degeneracies
(Figures~\ref{fig:6params_lensed}--\ref{fig:nu_mass}).  The latter concept
will likely become more important as we explore further data sets to constrain
cosmology within and beyond the 6-parameter $\Lambda$CDM paradigm.

Temperature and polarization anisotropies are correlated, and hence $TE$
contains information that enhances what is there from $TT$ and $EE$ alone.
A full measurement of $TT$, $EE$, and $TE$ yields one additional quantity per
pixel in the map (compared with just measuring $T$), and hence a total SNR that
is $\sqrt{2}$ times bigger than for $TT$ alone.  In addition to gaining back
the information lost due to the $TE$ correlation, when one measures all
three power spectra  one can also obtains additional information about how the
correlation itself changes (most easily seen in
Eq.~\ref{eq:decorrelatedston}--\ref{eq:decorrelatedstontheta}). In fact the
information gained from the correlation can sometimes be {\it greater\/} than
that from temperature (Ref.~\cite{Galli14}) or polarization
(Figure~\ref{fig:SNRparams}) alone.

Adding $B$-modes could in principle give one more quantity for each
pixel, although in practice the primordial signal is expected to be weak.
On the other hand CMB lensing provides an additional map of $\phi$, which
provides a whole other set of modes that can be used to constrain parameters.
For the standard $\Lambda$CDM model lensing helps to break the
$A_{\rm s}$--$\tau$ degeneracy, but for extensions to the standard model
(e.g., with neutrino mass included as an additional parameter) these data could
be even more useful in future.

Constraints from the CMB will continue to improve as we measure more modes from
polarization and from lensing.  There is certainly a bright near-term future
ahead as these measurements move towards being sample-variance-limited to small
angular scales.  In the longer term future there will be other secondary signals
extractable from CMB measurements, but ultimately, to dramatically increase the
number of modes probed, one will need to go to other observables (such as
redshifted 21-cm maps), which can provide 3D surveys of our past light cone.

\vskip 2cm

\appendix
\section{Decorrelating \textit{T} and \textit{E}}
\label{app:decorrelation}
As an alternative to the derivation of total CMB information content in
Section~\ref{sec:Fisher}, one can define new variables that are uncorrelated,
so that the covariance matrix becomes diagonal.
Doing this makes it clear that the improvement in SNR from including $E$-mode
polarization information is exactly $\sqrt{2}$, regardless of how $T$ and $E$
are correlated.  The change in adding $B$-modes is then trivial (since they are
uncorrelated with either $T$ or $E$). The approach we describe here is for any
two correlated data sets, the specific example will be temperature and
polarization data. Whether they are lensed or not does not change the
arguments, although the addition of lensing data themselves would require a
generalization of the method to deal with the associated correlations.

We can decorrelate temperature and polarization simply by rotating the data
$\{a^T_{\ell m}, a^E_{\ell m}\}$ into a new basis, designated as $\{t_{\ell m},
e_{\ell m}\}$, through an appropriate angle $\theta_{\ell}$:
\begin{align}
  \left(
  \begin{array}{c}
    t_{\ell m} \\
    e_{\ell m}
  \end{array} \right)
  &= \left(
        \begin{array}{cc}
        \phantom{-}\cos{\theta_{\ell}} & \sin{\theta_{\ell}} \\
        -\sin{\theta_{\ell}} & \cos{\theta_{\ell}}
        \end{array} \right)
  \left(
  \begin{array}{c}
    a^T_{\ell m} \\
    a^E_{\ell m}
  \end{array}
  \right).
\label{eq:rotation}
\end{align}
We denote the power spectra derived from this new set of variables as
$\lambda^{TT}_{\ell}, \lambda^{EE}_{\ell}, \lambda^{TE}_{\ell}$ and they are
given by
\begin{align}
  \left( \begin{array}{c}
   \lambda^{TT}_{\ell} \\
   \lambda^{EE}_{\ell} \\
   \lambda^{TE}_{\ell}
  \end{array} \right) &= \left( \begin{array}{ccc}
    \cos^2{\theta_{\ell}} & \sin^2{\theta_{\ell}} & \sin{2\theta_{\ell}} \\
    \sin^2{\theta_{\ell}} & \cos^2{\theta_{\ell}} & -\sin{2\theta_{\ell}} \\
    -\frac{1}{2}\sin{2\theta_{\ell}} & \frac{1}{2}\sin{2\theta_{\ell}} &
\cos{2\theta_{\ell}}
  \end{array} \right) \left( \begin{array}{c}
    C^{TT}_{\ell} \\
    C^{EE}_{\ell} \\
    C^{TE}_{\ell}
  \end{array} \right),
 \label{eq:Tmatrix}
\end{align}
or $\vec{\lambda}_{\ell} = R_{\ell}\vec{x}_{\ell}$, with $R_{\ell}$ defined as the
transformation above.
By demanding that $t_{\ell m}$ and $e_{\ell m}$ be uncorrelated (equivalently
that $\lambda^{TE}_{\ell} = 0$) we fix the angle to be
\begin{align}
  \theta_{\ell} &= \frac{1}{2}\tan^{-1}
 \left(\frac{2C^{TE}_{\ell}}{C^{TT}_{\ell} - C^{EE}_{\ell}}\right).
 \label{eq:theta}
\end{align}

Note that there are alternative approaches to decorrelating $T$ and $E$,
e.g., by leaving $T$ unaltered, while removing the correlated part from $E$
\cite{FE09}; the approach we describe here is easy to picture as a rotation.
The covariance matrix for these new power spectra can simply be derived by
computing the 4-point functions of $t$ and $e$. However,
a much simpler method is to
take the previous covariance matrix (i.e., $\mathbb{C}$) and make the following
replacements: $C^{TT} \rightarrow \lambda^{TT}$; $C^{EE} \rightarrow
\lambda^{EE}$; and $C^{TE} \rightarrow 0$. The covariance and inverse
covariance matrices then become
\begin{align}
  \mathbb{L}_{\ell} &= \frac{2}{(2\ell+1)f_{\rm sky}}
    \left(
    \begin{array}{ccc}
      {\lambda^{TT}_{\ell}}^2 & 0 & 0 \\
      0 & {\lambda^{EE}_{\ell}}^2 & 0 \\
      0 & 0 & \frac{1}{2}\lambda^{TT}_{\ell}\lambda^{EE}_{\ell}
    \end{array} \right),
  \label{eq:lambdacov}\\
  {\mathbb{L}}^{-1}_{\ell} &= \left(\ell + \frac{1}{2}\right)f_{\rm sky}
    \left(
    \begin{array}{ccc}
      1/{\lambda^{TT}_{\ell}}^2 & 0 & 0 \\
      0 & 1/{\lambda^{EE}_{\ell}}^2 & 0 \\
      0 & 0 & 2/\lambda^{TT}_{\ell}\lambda^{EE}_{\ell}
    \end{array} \right),
\end{align}
respectively (or equivalently $\mathbb{L}_{\ell} =
R_{\ell}\mathbb{C}_{\ell}R_{\ell}^{\sf T}$ and $\mathbb{L}_{\ell}^{-1} =
(R_{\ell}^{\sf T})^{-1}\mathbb{C}_{\ell}^{-1}R_{\ell}^{-1}$).
The data vector takes the simple form
\begin{align}
  \vec{\lambda}_{\ell} &= \left( \begin{array}{c}
    \lambda^{TT}_{\ell} \\
    \lambda^{EE}_{\ell} \\
    \lambda^{TE}_{\ell}
  \end{array} \right).
  \label{eq:datavec}
\end{align}
Note that even although $\lambda^{TE}_{\ell}$ is zero, this does not imply that
for a general parameter $p$, $\partial{\lambda^{TE}_{\ell}}/\partial p$ will
vanish.

The transformation performed here leaves the Fisher matrix unchanged.  This
is because
\begin{align}
  F'_{ij} &= \sum_{\ell} \frac{\partial \vec{\lambda}_{\ell}^{\sf T}}
 {\partial p_i} \mathbb{L}_{\ell}^{-1}
 \frac{\partial \vec{\lambda}_{\ell}}{\partial p_j} \\
    &= \sum_{\ell} \frac{\partial \vec{x}^{\sf T}_{\ell}}
 {\partial p_i} R^{\sf T}_{\ell} (R_{\ell}^{\sf T})^{-1}
 \mathbb{C}_{\ell}^{-1}R_{\ell}^{-1}R_{\ell}
 \frac{\partial \vec{x}_{\ell}}{\partial p_j} \\
    &= \sum_{\ell} \frac{\partial \vec{x}^{\sf T}_{\ell}}
 {\partial p_i}\mathbb{C}_{\ell}^{-1}
 \frac{\partial \vec{x}_{\ell}}{\partial p_j} \\
    &= F_{ij}.
  \label{eq:fisherinvariant}
\end{align}
Hence, for a single scaling parameter $p$, such that $\partial
\vec{\lambda}_{\ell}/\partial p = \vec{\lambda}_{\ell}/p$, we trivially find
\begin{align}
  F_{pp} &= \frac{2}{p^2}f_{\rm sky}\sum_{\ell}
 \left(\ell + \frac{1}{2}\right),
  \label{eq:fisherdiag}
\end{align}
which is the same as the result of Eq.~(\ref{eq:finalFisher}).

We can also consider a general parameter $p$ like in
Section~\ref{sec:fullfisherinformation}, in which
case we will have a fixed rotation matrix $R$ and hence
\begin{equation}
  \left(\frac{S}{N}\right)^2 = \sum_{\ell} \mathcal{N} \left[
  \left(\frac{d\ln \lambda^{TT}_{\ell}}{d\ln p}\right)^2 + \left(\frac{d\ln
  \lambda^{EE}_{\ell}}{d\ln p}\right)^2
   + \frac{2}{\lambda^{TT}_{\ell}\lambda^{EE}_{\ell}} \left(\frac{d
  \lambda^{TE}_{\ell}}{d\ln p}\right)^2\right].
  \label{eq:decorrelatedston}
\end{equation}
The interpretation of the terms on the right-hand side is quite clear:
the first two are the information coming from temperature and polarization
anisotropies, respectively, after removing their correlation;
and the final term is the information coming from
the correlation itself. This can be seen by re-writing this in terms of
$C^{TE}$ and $\theta_\ell$ as
\begin{align}
  \left(\frac{S}{N}\right)^2 &= \sum_{\ell} \mathcal{N} \left[
  \left(\frac{d\ln \lambda^{TT}_{\ell}}{d\ln p}\right)^2 + \left(\frac{d\ln
  \lambda^{EE}_{\ell}}{d\ln p}\right)^2 \right. \notag\\
  &\qquad\qquad \left. + \frac{\left(C^{TT}_{\ell}
   - C^{EE}_{\ell}\right)^2}{2\lambda^{TT}_{\ell}\lambda^{EE}_{\ell}}
  \left(\cos{2\theta_{\ell}}
   \frac{d \tan{2\theta_{\ell}}}{d\ln p}\right)^2\right].
  \label{eq:decorrelatedstontheta}
\end{align}
So we see, as we found in Section~\ref{sec:fullfisherinformation}, that
depending on the parameter $p$, there is extra information to be found from
the correlation itself, even if $C^{TE}_{\ell} = 0$.

\section{The fully correlated case}
\label{app:blensing}
Note that the Fisher formalism we have employed in this paper relies on the
covariance matrix being invertible. In a realistic experiment this is not an
issue, because, regardless of the true correlation between modes, noise will
regularize the problem, such that the covariance will always be invertible.
Additionally, on physical grounds, it is clear that we need not worry
even in the case of no noise, because we will {\it never\/} be in the
situation where $r^2_{\ell} = 1$. We can see the problem with $r^2_\ell\to1$
in the following way -- adding $E$ should double the number of independent
modes, irrespective of how correlated $T$ and $E$ are, {\it except\/} for the
case $r_\ell^2=1$, when there is no new information from $E$ (and hence the
mode count appears to be discontinuous as $r^2_\ell\to1$).  Clearly this
situation is hypothetical and hence should not unduly concern us.

However, there is a similar case where we {\it would\/} have a singular
covariance and that is the case of $E$-modes, lensing, and lensed $B$-modes.
It is apparent that the lensed $B$-modes are completely determined by $E$-modes
and lensing modes, $\phi$. Therefore we can simply ask what there is to be
gained from actually measuring the $B$-modes. The answer is that (in principle)
there is nothing gained in this measurement! This is true if we assume that a
perfect measurement of $E$, $B$, and $\phi$ can be performed. From an
experimental
point of view this will never really be true, due to the presence of noise
(i.e., one can always hope to beat down the noise in lensed $B$-modes by
actually measuring them). Of course the Fisher-formalism assumes the model to
be correct, so from a theoretical point of view (and looking beyond standard
$\Lambda$CDM), measuring lensed $B$-modes will of course always have merit.

\acknowledgments
This research was supported by the Natural Sciences and Engineering Research
Council of Canada and by the Canadian Space Agency.  We thank Jim Zibin for
several enjoyable discussions on this topic and for providing useful comments
on the paper.

\bibliographystyle{apsrev4-1}
\bibliography{cosmo_2D}

\end{document}